\documentstyle[12pt,./psfig,./aaspp4]{article}

\newcommand{\Wo}{\mbox{${\rm W}_0$}}
\newcommand{\msun}{\mbox{${\rm M}_\odot$}}
\newcommand{\msunyr}{ \mbox{ ${\rm M}_\odot \,{\rm yr}^{-1}$ } }
\newcommand{\lsun}{\mbox{${\rm L}_\odot$}}

\newcommand{\kms}{\mbox{${\rm km~s}^{-1}$}}
\newcommand{\kmpspc}{\mbox{${\rm km~s}^{-1} \,{\rm pc}^{-1}$}}

\newcommand{\paperIII}{\mbox{paper~{\rm III}}}
\newcommand{\PaperIII}{\mbox{Paper~{\rm III}}}
\newcommand{\paperIV}{\mbox{paper~{\rm IV}}}
\newcommand{\nbody}{\mbox{{{\em N}-body}}}
\newcommand{\thm}{\mbox{${t_{\rm hm}}$}}

\newcommand{\thc}{\mbox{${t_{\rm hm}}$}}
\newcommand{\tvir}{\mbox{${t_{\rm vir}}$}}

\newcommand{\trlx}{\mbox{${t_{\rm rlx}}$}}
\newcommand{\trxh}{\mbox{${t_{\rm hrx}}$}}
\newcommand{\trxt}{\mbox{${t_{\rm trx}}$}}

\newcommand{\mgal}{\mbox{${M_{\rm Gal}}$}}

\newcommand{\rcore}{\mbox{${r_{\rm core}}$}}

\newcommand{\rvir}{\mbox{${r_{\rm vir}}$}}

\newcommand{\rhm}{\mbox{${r_{\rm hm}}$}}
\newcommand{\rgc}{\mbox{${r_{\rm GC}}$}}
\newcommand{\rgal}{\mbox{${r_{\rm GC}}$}}

\newcommand{\rtide}{\mbox{${r_{\rm tide}}$}}

\newcommand{\rLf}{\mbox{${r_{\rm L1}}$}}

\newcommand{\FP}{\mbox{\rm Fokker-Planck}}

\newcommand{\tend}{\mbox{$t_{\rm end}$}}
\newcommand{\pc}{\mbox{${\rm pc}$}}

\def\unit#1{{\mbox{[{\rm #1}]}}}
\def\apgt{\ {\raise-.5ex\hbox{$\buildrel>\over\sim$}}\ }
\def\aplt{\ {\raise-.5ex\hbox{$\buildrel<\over\sim$}}\ }

\begin{document}


\title{The lives and deaths of star clusters near the Galactic center}

\author{Simon F.\ Portegies Zwart\altaffilmark{1}$^\star$,
	Junichiro Makino\altaffilmark{2},
	Stephen L.\ W.\ McMillan\altaffilmark{3},
    	Piet Hut\altaffilmark{4}
}

$^\star$ SPZ is a Hubble Fellow \\

\altaffiltext{1}{Massachusetts Institute of Technology, Cambridge, MA 02139, USA}
\altaffiltext{2}{Department of Astronomy, University of Tokyo, 7-3-1 Hongo,
	       Bunkyo-ku,Tokyo 113-0033, Japan}
\altaffiltext{3}{Department of Physics,
		 Drexel University,
                 Philadelphia, PA 19104, USA}
\altaffiltext{4}{Institute for Advanced Study,
		 Princeton, NJ 08540, USA}

\lefthead{Portegies Zwart et al.}
\righthead{Dissection of an open star cluster}

\begin{abstract}
We study the evolution and observability of young, compact star
clusters near the Galactic center, such as the Arches and Quintuplet
systems.  The clusters are modeled by integrating the equations of
motion of all stars while accounting for the internal evolution of
stars and binaries, as well as the effect of the Galactic tidal field.
We find that clusters within 150\,pc of the Galactic center dissolve
within $\sim55$ Myr, but their projected densities drop below the
background density in the direction of the Galactic center within only
a few Myr, effectively making these clusters undetectable after that
time.
Detailed observations of the Arches cluster, taken at face value,
suggest that its mass function is unusually flat and that the cluster
contains an overabundance of stars more massive than 20\,\msun.  Our
dynamical analysis, however, shows that the observed characteristics
of the Arches cluster are consistent with a perfectly normal initial
mass function.  The observed anomalies are then caused by a
combination of observational selection effects and the dynamical
evolution of the cluster.  We calibrate the current parameters of the
Arches cluster using a normal initial mass function and conclude that
the cluster is more massive than 40\,000\,\msun, has a half mass
radius of about 0.35\,pc and is located between 50 and 90\,pc from the
Galactic center.
\keywords{binaries: close ---
 	 stars: blue stragglers ---
	 stars: evolution ---
	 globular clusters: general ---
	 globular clusters: 30 Doradus, Arches, Quintuplet --
	}  

\end{abstract}

\section{Introduction}
A number of young, dense star clusters have been observed within the
inner few hundred parsecs of the Galactic center.  Best known are the
Arches cluster (Object 17, Nagata et al.\,
1995)\nocite{1995AJ....109.1676N} and the Quintuplet cluster
(AFGL\,2004, Nagata et al.\, 1990; Okuda et al.\,
1990).\nocite{1990ApJ...351...83N}\nocite{1990ApJ...351...89O}
However, it is likely that others exist, as these clusters lie behind
thick layers of obscuring material (Portegies Zwart et al.\,
2001a).\nocite{2001ApJ...546L.101P} Most are expected to be invisible
at optical wavelengths, but should be readily detectable in the
infrared (Vrba et al.\, 2000) or in the 2MASS survey Dutra \& Bica
2000).\nocite{2000A&A...359L...9D}

The Arches and the Quintuplet clusters are the Galactic counterparts
of R\,136, the central star cluster in NGC\,2070: the 30\,Doradus
region in the Large Magellanic Cloud (Massey \& Hunter
1998)\nocite{1998ApJ...493..180M}.  The structural parameters of these
clusters---masses, radii, and density profiles--- are quite similar,
as are their ages.  R\,136, however, is located far from the
perturbing influence of the Galaxy and the tidal effect of the LMC is
negligible (Portegies Zwart et al.\ 1999).  The Arches and the
Quintuplet clusters, on the other hand, lie at projected distances of
$\aplt 40$\,pc of the Galactic center, and their evolution is strongly
affected by the Galactic tidal field.

We study the evolution of clusters like Arches and Quintuplet using a
fully self-consistent star-cluster model in which the dynamics of
stars are followed by direct {\nbody} integration and the evolution
of individual stars and binaries are followed using a stellar and
binary evolution program.  The importance of the tidal field of the
Galaxy is studied by repeating each model calculations at several
distances from the Galactic center. We compare the results of our
calculations with detailed observations of the Arches cluster and with
model calculations performed by others.

We find that, while all our models start with a mass function
comparable to that in the solar neighborhood, to an observer the
current mass function may well appear to be much flatter.  The
unusually flat mass functions observed in both the Arches and the
Quintuplet clusters may therefore in part be attributed to a
combination of observational selection effect, the age of the cluster
and the location at which the measurements were taken.

The star clusters studied in this paper experience core collapse
within a few million years. During this phase and at later time many
stars experience collisions with other stars. The clusters finally
dissolve in the tidal field of the Galaxy. The disruption of the
cluster is driven mostly by tidal stripping and two-body
relaxation. Stellar evolution plays only a minor role in the
disruption of the clusters.

The clusters expand as they age, causing their surface densities to
decline.  Clusters older than about 5\,Myr often have surface
densities below that of their surroundings, making these clusters
virtually undetectable.  Since the modeled clusters disrupt at ages
greater than 5 Myr, we estimate that there may be many clusters like
the Arches and Quintuplet, but only the youngest are observable.  This
suggests that the formation of Arches and Quintuplet-like clusters may
be a continuous process.  We cannot exclude, however, that these
clusters were formed in a recent burst of star formation and that
older counterparts do not exist.

The numerical methods and selection of the initial conditions are
discussed in Sec.\,2.  The results are presented in Sec.\,3 and
discussed in Sec.\,4.  We summarize and conclude in Sec.\,5.

\section{The model}\label{Sect:Units}
The {\nbody} portion of the simulations is carried out using the {\tt
kira} integrator operating within the ``Starlab'' software
environment.  Time integration of stellar orbits is accomplished using
a fourth-order Hermite scheme (Makino \& Aarseth
1992).\nocite{1992PASJ...44..141M} {\tt Kira} also incorporates block
time steps (McMillan 1986ab; Makino
1991)\nocite{1986ApJ...307..126M}\nocite{1986ApJ...306..552M}
\nocite{1991ApJ...369..200M} special treatment of close two-body and
multiple encounters of arbitrary complexity, and a robust treatment of
stellar and binary evolution and stellar collisions.  The
special-purpose GRAPE-4 (Makino et al.\
1997)\nocite{1997ApJ...480..432M} system is used to accelerate the
computation of gravitational forces between stars.  The treatment of
stellar mass loss is described by Portegies Zwart et
al.~(1998).\nocite{1998A&A...337..363P} A concise description of the
Starlab environment is given by Portegies Zwart et al.~(2001b,
\paperIV);\nocite{2001MNRAS.321..199P} additional information is
available at {\tt http://manybody.org}.

Evolution of stars and binaries is handled using the prescription
given by Portegies Zwart \& Verbunt (1996, Sec.\,2.1).\nocite{pzv96}
However, some changes are made to the treatment of main-sequence mass
loss in massive stars (see Portegies Zwart et al.\, 1999, hereafter
\paperIII),\nocite{pzmmh99} and we incorporate a more extended
prescription for stellar collisions, as described by Portegies Zwart,
Hut \& Verbunt\, (1997).\nocite{pzhv97}

The system of units used internally in the $N$-body models is defined
by $M = G = -4E = 1$, where $E$ is the initial internal energy of the
stellar system, $M$ is the total mass in stars and $G$ is the
gravitational constant (Heggie \& Mathieu 1986).\nocite{HM1986} 

The time required for a star to cross the clusters' virial radius
\rvir\ is
\begin{equation}
	\tvir = \left( {GM \over \rvir^3} \right)^{-1/2}.
\label{eq:virial_crossing_time}\end{equation}

For the runs considered here, the {\nbody} and physical length scales
are connected by the requirement that initially the star cluster is in
virial equilibrium and exactly fills its zero-velocity surface in the
Galactic tidal field (see Sec.~\ref{sect:tidal} below).  Once the
cluster mass and distance to the Galactic center are known, the
clusters' size scale is determined.  The half mass crossing time
($\thc = 2\sqrt{2}\tvir$) can then be expressed in convenient physical
units as
\begin{equation}
	\thc \simeq 42.2 \left( {[\msun] \over M} \right)^{1/2}
		  \left( {\rhm \over [{\rm pc}]}\right)^{3/2} \, \unit{Myr}.
\label{eq:crossing_time}\end{equation}
For most density profiles the half-mass radius {\rhm} is slightly
($\aplt 25$\%) smaller than \rvir.

The half-mass relaxation time is calculated as (Spitzer
1987):\nocite{1987degc.book.....S}
\begin{equation}
	\trxh = \left( {\rhm^3 \over G M} \right)^{1/2} 
		{N \over 8 \log\Lambda}.
\label{Eq:trlx}\end{equation}
Here $\Lambda \equiv \gamma N$ is the coulomb logarithm;
$\gamma\sim0.4$ is a scaling factor introduced to model the effects of
the cut-off in the long range Coulomb logarithm (see Giersz \& Heggie
1996; 1994).\nocite{1996MNRAS.279.1037G}\nocite{1994MNRAS.268..257G}
We will also use the relaxation time at the tidal radius \trxt, for
which we use Eq.\,\ref{Eq:trlx} but with the tidal radius \rtide\,
substituted for \rhm.

\subsection{Selection of initial conditions}

The two clusters on which we concentrate here, the Arches and the
Quintuplet, lie in projection within $\sim 40$\,pc of the Galactic
center.  Table\,\ref{Tab:observed} lists some observed parameters of
these clusters, along with some other clusters (and cluster
candidates) having comparable characteristics.  All have masses of
$\sim 10^4\,\msun$, are very compact, $\rhm\aplt 1$\,pc, and are only a
few million years old.  However, of the systems listed, only the
Arches and Quintuplet are significantly perturbed by external tidal
fields.

\begin{table*}[htbp!]
\caption[]{Observed parameters for some young, dense clusters.
Columns give the cluster name, reference, age, mass, distance to the
Galactic center (the cluster R\,136 is located in the Large Machelanic
Cloud), the tidal radius \rtide\, and the half-mass radius \rhm.  The
last two columns give the density within the half-mass radius and the
half mass relaxation time. \\ }
\begin{flushleft}
\begin{tabular}{ll|rrrrrcc} \hline
Name   &ref& Age  &   M     & \rgc & \rtide & \rhm & $\rho_{\rm hm}$&\trxh\\ 
       &&[Myr]& [$10^3$\,\msun] & \multicolumn{3}{c}{--------- [pc]
---------} &
			 [$10^5$\,\msun/pc$^3$] & [Myr] \\ \hline
R\,136    &a& 2--4 & 21--79   & LMC  &$\apgt20$& $\sim 0.5$  &0.4--1.5 & 70 \\
Arches    &b& 1--2 & 12--50   & 30   & 1       & 0.2 &3.6--15 & 12 \\ 
Quintuplet&c& 3--5 & 10--16   & 35   & 1       & $\sim 0.5$& 0.14--0.31 & 12 \\
NGC\,3603 &d& 2--3 &  3--13   &few\,k& 4--5    & 0.23      &0.59--2.6 & 44 \\
W43 (=G30.8-0.2)&e&$\aplt10$& ?$^\star$ \\
Norma     &f&     &?$^\circ$ & few\,k & & \\
{\em nameless}&g&$\aplt10$  & ?$^\bullet$ 
				 &12-15k& $>1$  &0.2--0.4& 0.45--4.5\\
\hline
\end{tabular} \\
\smallskip
References:
a) Brandl et al.\, (1996);\nocite{1996ApJ...466..254B} 
   Campbell et al.\, (1992);\nocite{1992AJ....104.1721C}
   Massey \& Hunter (1998).\nocite{1998ApJ...493..180M}
b) Figer et al.\, (1999a);\nocite{1999ApJ...525..750F}
c) Glass, Catchpole \& Whitelock (1987);\nocite{1987MNRAS.227..373G} 
   Nagata et al.\, (1990);\nocite{1990ApJ...351...83N}
   Figer, Mclean \& Morris (1999b).\nocite{1999ApJ...514..202F} 
d) Brandl (1999)\nocite{1999A&A...352L..69B}
e) Blum, Daminelli \& Conti (1999)\nocite{1999AJ....117.1392B}
f) Moffat (1976)\nocite{1976A&A....50..429M}
g) Vrba et al.\, (2000)\nocite{2000ApJ...533L..17V} \\
\smallskip
$^\star$ Contains approximately 100 spectral type O and WN stars but
further information is not available.\\ 
$^\circ$ Contains a dozen embedded early-type stars 
surrounding the M3\,Ia star HD\,143183. \\
$^\bullet$ Contains at least
13 early type O stars, the cluster does not have a name but is located
near the soft gamma-ray repeater SGR 1900+14 (also SGR 1806-20 may be
associated with a young and compact star cluster [van Kerkwijk et
al.\, 1995; Kulkarni et al.\,
1995]).\nocite{1995ApJ...444L..33V}\nocite{1995ApJ...440L..61K} \\
\end{flushleft}
\label{Tab:observed} 
\end{table*}

\subsection{The tidal field near the Galactic center}
\label{sect:tidal}

A star cluster embedded in the Galactic tidal field is not spherically
symmetric.  Rather, it is flattened and the stellar velocity
distribution is anisotropic, particularly in the outer regions.  The
initial models which best describe such a cluster are the anisotropic
density profiles described by Heggie \& Ramamani
(1995).\nocite{1995MNRAS.272..317H} As in the usual spherically
symmetric King (1966)\nocite{1966AJ.....71...64K} models, the density
profile is described by the dimensionless parameter \Wo.  Higher
values of {\Wo} indicate a more centrally concentrated cluster (see
the illustration in Fig.\,\ref{fig:SPZ_W147}).

Following Heggie \& Ramamani, we model the tidal potential $\phi_T$ by
the quadrupole formula
\begin{equation}
	\phi_T(x, y, z) = -{\textstyle\frac12} (\alpha_1 x^2 + \alpha_3 z^2)\,.
\end{equation}
The cluster is taken to move on a circular orbit around the Galactic
center.  The $x$-axis in the rotating frame of reference centered on
the cluster always points toward the Galactic center; the $z$-axis is
perpendicular to the orbital plane.  The quantities $\alpha_1 (>0)$
and $\alpha_3 (<0)$ are conveniently expressed in terms of the
kinematic Oort constants $A$ and $B$ (Oort 1927\nocite{Oort1927}) and
the local Galactic density $\rho_G$ as:
\begin{eqnarray}
	\alpha_1 &=& -4A(A-B)\,, \nonumber\\
	\alpha_3 &=& 4\pi G\rho_G + 2(A^2-B^2)\,.
\end{eqnarray}
Loosely speaking, we can think of $\alpha_1$ as determining the
overall strength of the tidal field, while the ratio
$\alpha_1/\alpha_3$ determines its geometry.

The Oort constants $A$ and $B$ are defined as
\begin{eqnarray}
	A &=& {1\over 2} \left(  {v_c \over r}
		- {dv_c \over dr} \right),  \nonumber \\
	B &=& -{1\over 2} \left( {v_c \over r}
		+ {dv_c \over dr} \right)\,,
\label{Eq:OortAB}\end{eqnarray}
where $v_c$ is the circular rotational velocity:
\begin{equation}
	v_c = \sqrt{ {GM_{\rm Gal}(r) \over r} }.
\end{equation}
Taking the mass of the Galaxy within the cluster's orbit at distance
{\rgc} ($\aplt100$\,pc) from the Galactic center to be (Mezger et
al.\, 1999)\nocite{1999A&A...348..457M}
\begin{equation}
	\mgal(\rgc) = 4.25\times10^6 \left({\rgc \over \unit{pc}}
		     	           \right)^{1.2}           \;\;\msun,
\label{Eq:Mgal}\end{equation}
we find
\begin{eqnarray}
	 v_c     &=& 136.8 \left( {r \over
                        \unit{pc}} \right)^{0.1}  \;\;  
						\kms \nonumber \\ 
	{dv_c \over dr}  &=&  13.7 \left( {r
					\over 
			\unit{pc}} \right) ^{-0.9}  \;\; \kmpspc\,.
\label{Eq:vc}\end{eqnarray}
and hence
\begin{eqnarray}
	A	&\simeq&  61.5\  \rgal^{-0.9} \;\;\kmpspc \nonumber \\
	B	&\simeq& -75.3\  \rgal^{-0.9} \;\;\kmpspc \nonumber \\
	\rho_G	&\simeq& 4.06\  \times10^5 \rgal^{-1.8}
						\;\;\msun\,{\rm pc}^{-3}\,.
\label{Eq:rhogal}
\end{eqnarray}
Table\,\ref{Tab:tidalfield} lists these parameters at selected
Galactocentric distances.

\begin{table*}[htbp!]
\caption[]{ Parameters for the Galactic tidal field of the calculated
star clusters at selected distances from the Galactic center.  Each
row gives the distance to the Galactic center, the local stellar
density, and the Oort $A$ and $B$ constants.  The first row gives that
value for the local stellar density from Hill, Hiditch \& Barnes
(1985)\nocite{1979MNRAS.186..813H} and the Oort constants in the solar
vicinity from Olling \& Merrifield (1998)\nocite{1998MNRAS.297..943O}.
}
\begin{flushleft}
\begin{tabular}{lrccc} \hline
$r_{\rm GC}$ &$\rho_G(r)$      & $v_c$  & $A$ & $B$ \\ 
\unit{pc}     & [\msun/pc$^3$]& [\kms] & [\kmpspc] & [\kmpspc] \\
\hline
Sun &0.11&$184\pm8$&$11.3\pm1.1 \times 10^{-3}$&$-13.9\pm0.9 \times 10^{-3}$\\
 34 &711  &  193       & 2.55 & -3.12 \\ 
 90 &123  &  212       & 1.06 & -1.30 \\
150 & 49  &  224       & 0.67 & -0.82 \\ 
\hline
\end{tabular}
\end{flushleft}
\label{Tab:tidalfield} \end{table*}

The distance from the center of the star cluster (of mass $M$) to the
first Lagrange point (the Jacobi radius) is
\begin{equation}
	\rLf \equiv \left( \frac{-M}{\alpha_1} \right)^{1/3}
\label{Eq:L1}\end{equation}
Substitution of Eq.\,\ref{Eq:Mgal} into Eq.\,\ref{Eq:L1} yields
\begin{equation}
	\rLf \simeq 4.90\times10^{-3} \left( {M \over [\msun]}
	                     \right)^{1/3} 
			     \left( {\rgc \over [{\rm pc}]}
			     \right)^{0.6}\;\; {\rm pc}.
\end{equation}

\subsection{Initial cluster structure}\label{sect:initial}

We adopt initial models comparable to those observed for the Arches and
the Quintuplet clusters.  By varying the density profile and the
distance to the Galactic center we study how the cluster evolution
depends on the initial conditions.

Our calculations start with 12k (12288) stars at zero age.  We assign
stellar masses $m$ in the range $0.1\,\msun < m < 100\,\msun$ from the
mass function suggested for the Solar neighborhood by Scalo
(1986).\nocite{scalo86} The median mass of this mass function is about
0.3\,\msun; the mean mass is $\langle m \rangle \simeq 0.6\,\msun$.
For models with 12k stars this results in a total cluster mass of
$\sim 7500$\,\msun.  We adopt a mass function which is applicable to
the Solar neighborhood rather than the flatter mass spectrum suggested
by observations (Figer et al., 1999a) in order to determine whether the
flat mass spectrum can be attributed to cluster dynamical evolution.
Initially all stars are single, although binaries do form via
three-body encounters, in which one star carries away the excess
energy and angular momentum necessary for two other stars to become
bound.

We adopt three standard distances from the Galactic center: 34\,pc,
90\,pc and 150\,pc. The shape of the zero-velocity surface in the
tidal field of the Galaxy only depends on $\alpha_3/\alpha_1$, which
is independent of $\rgal$.  However, this does not mean that
the models can be scaled with respect to distance from the Galactic
center using the relaxation time alone. The time scale on which the
stars in the cluster evolve depends on the size of the cluster via
Eq.\,\ref{eq:crossing_time}; when the cluster is located further from
the Galactic center the crossing time increases and the stars evolve
relatively more quickly compared to the dynamical evolution of the
cluster. This results in a more active mass loss by stellar evolution
for clusters which are farther from the Galactic center.

\begin{figure}[htbp!]
\hspace*{1cm}
\psfig{figure=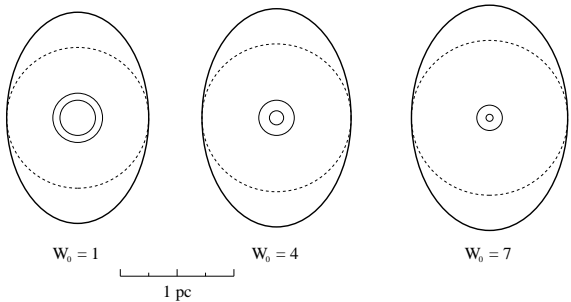,width=7.5cm,angle=-90}
\caption[]{Illustration of the structure of star clusters with \Wo=1
(left), \Wo=4 (middle) and \Wo=7 (right).  The outer ellipse
represents the zero velocity surface of the cluster in the Galactic field.  The
two inner circles represent the half-mass and core radii.  The dashed
circle shows the `tidal radius' that would be obtained if the initial
model were described by King's spherically symmetric density profile.
The Galactic center is to the top of the figure, at a distance of
150\,pc. The length scale is indicated by the horizontal bar.
}
\label{fig:SPZ_W147}
\end{figure}

Initial density profiles and velocity dispersions are taken from
Heggie \& Ramamani (1995)\nocite{1995MNRAS.272..317H} with $\Wo
= 1$, $\Wo = 4$ and $\Wo = 7$, for a total of 9 models.  At birth the
clusters are in virial equilibrium and exactly fill their critical
zero-velocity surfaces (``Roche lobes'') in the Galactic tidal field.
Tab.\,\ref{Tab:N12kinit} presents a summary of the adopted initial
models.  Figure\,\ref{fig:SPZ_W147} indicates the shape and structure
of models with \Wo=1, 4 and 7 at $\rgal = 150$ pc; models closer to
the Galactic center are identical in shape, but the scale is
different.  The zero-velocity surfaces of the various models are
represented as ellipses.

We test the reproducibility of our calculations by performing several
calculations per set of selected initial conditions.  Each calculation
was performed twice with a different random seed. In addition, we performed a
set of calculations without stellar evolution for models with $\Wo=1$,
4 and 7. These same initial realizations were rerun with stellar
evolution switched on, for $\rgc=34$\,pc, 90\,pc and 150\,pc.  In total,
30 simulations were performed.

\begin{table*}[htbp!]
\caption[]{Overview of initial conditions for the simulations
performed.  Each row lists the model name, the distance to the
Galactic center, the initial King parameter \Wo, the initial tidal
relaxation time, the initial half-mass relaxation time (see
Eq.\,\ref{Eq:trlx} and see paper IV for the version in more usual
astronomical units), the half-mass crossing time (see
Eq.\,\ref{eq:crossing_time}), and the initial core radius, virial
radius, and the tidal radius.  The final four columns give the number
of stellar collisions observed in each calculation, the time of core
collapse, the time at which the cluster mass dropped below 5\% of the
initial mass (about 375\,\msun) and the collision rate in number of
collisions per million years.}

\begin{flushleft}
\begin{tabular}{lrc|rrclcr|crrrr} \hline
Model  &$r_{\rm GC}$
            &\Wo&\trxt&\trxh& \thm&\rcore&\rvir & \rtide     &$n_{\rm coll}$
						      &$t_{cc}$ 
						      &$t_{\rm end}$ 
					      &$n_{\rm coll}/t_{\rm end}$ \\
       &[pc]&   &\multicolumn{2}{c}{--- [Myr] ---}&[Kyr]
	&\multicolumn{3}{c|}{--------- [pc] ---------} &
			    &\multicolumn{2}{c}{--- [Myr] ---} & [Myr$^{-1}$] \\ 
\hline
R34W1 & 34& 1& 53&  5.5 & 43& 0.12 & 0.20 & 0.76 & 5&1.9&  7.4& 0.68 \\
R34W4 & 34& 4& 53&  3.2 & 27& 0.05 & 0.14 & 0.76 & 8&1.2& 12.7& 0.63 \\
R34W7 & 34& 7& 53&  0.36&  9& 0.018& 0.077& 0.80 &24&0.4& 12.0& 2.00 \\ 
\hline    		       
R90W1 & 90& 1&134& 14.6 &105& 0.19 & 0.36& 1.4  & 7&1.9& 25.6& 0.27 \\
R90W4 & 90& 4&134&  8.1 & 68& 0.09 & 0.26& 1.4  &15&1.2& 32.6& 4.02 \\ 
R90W7 & 90& 7&134&  1.0 & 25& 0.032& 0.14& 1.4  &14&1.4& 32.4& 32.4 \\
\hline    		       
R150W1&150& 1&218& 23.6 &169& 0.30 & 0.50& 1.9 & 5&2.0& 45.8& 1.94 \\ 
R150W4&150& 4&218& 13   &110& 0.14 & 0.36& 1.9 & 8&2.0& 53.4& 4.08 \\ 
R150W7&150& 7&218&  4.5 & 40& 0.044& 0.19& 2.0 &14&0.4& 55.0& 12.2  \\
\hline     		       
\end{tabular}
\end{flushleft}
\label{Tab:N12kinit} \end{table*}

Stars are removed from the {\nbody} system when their distance from
the center of the cluster exceeds 3\rLf. In {\nbody} calculations
it is not always trivial to determine the moment the cluster
dissolves, as a few stars may remain bound for an extended period of
time (Portegies Zwart et al 1998).\nocite{1998A&A...337..363P} In our
models we identify the cluster's disruption as the moment when no
stars lie within the zero velocity surface.  Typically, a few hundred
stars remain in the {\nbody} system (within $3\rLf$) at this time.

\section{Results}\label{Sect:results}\label{sect:results}

We first discuss the global parameters of all models.  Later (in
Secs.\,\ref{sect:mass_function} and \ref{sect:density_profile}) we
will concentrate on a few representative models and compare these with
the observed clusters.

\subsection{The evolution of the cluster}

\begin{figure}[htbp!]
(a)\psfig{figure=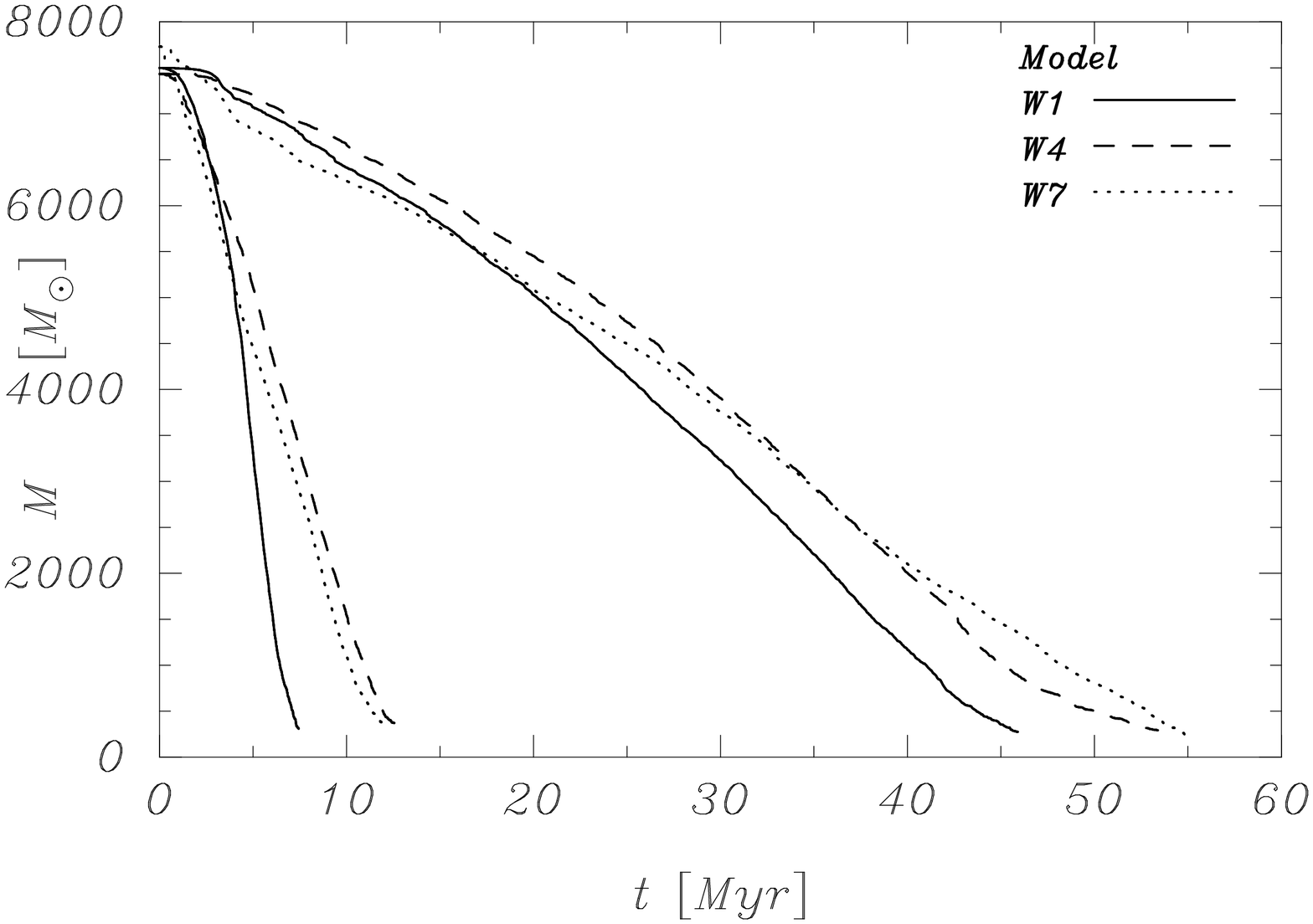,width=7.5cm,angle=-90}
(b)\psfig{figure=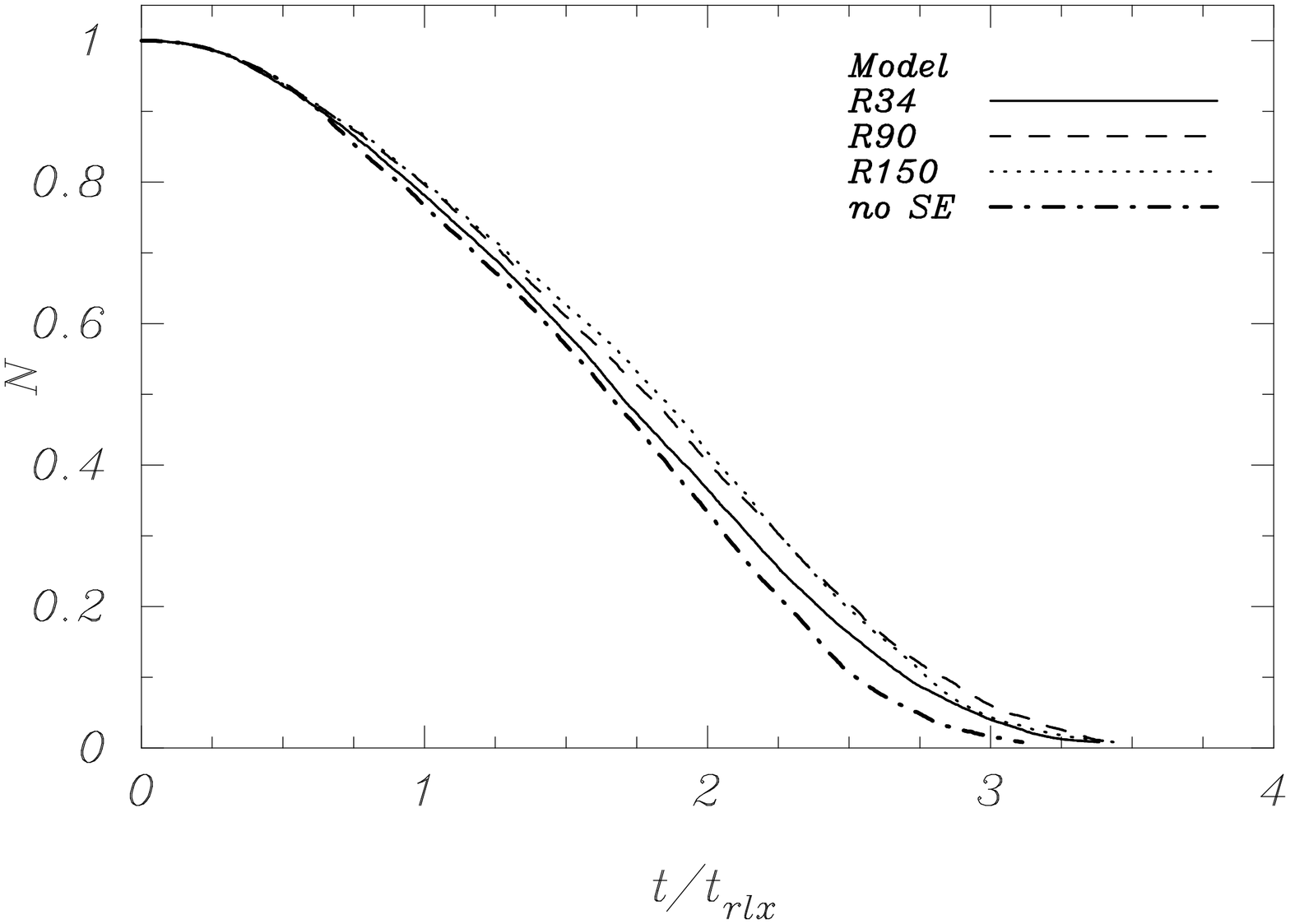,width=7.5cm,angle=-90}
\caption[]{(a) Evolution of total cluster mass for models with \Wo=1
(solid lines), \Wo=4 (dashed lines) and \Wo=7 (dots) at distances of
34\,pc (left lines) and 150\,pc (right) from the Galactic center.
(b) The evolution of the number of stars (renormalized to the initial
total of 12k) for the models with \Wo=4 as a function of time in units
of the initial half-mass relaxation time (see Eq.\,\ref{Eq:trlx2}) for
$\rgc=34$ (solid), $\rgc=90$ (dashes) and $\rgc=150$ (dots).  The
dash-dotted line gives the evolution of the number of stars for the
model without stellar evolution.  These models were all calculated
using the same realization of the initial conditions.  }
\label{fig:SPZ_W147N12k_TM}
\end{figure}

Figure\,\ref{fig:SPZ_W147N12k_TM}(a) shows the mass evolution of the
models listed in Table \ref{Tab:N12kinit}.  Not surprisingly, clusters
located at larger distances from the Galactic center tend to live
longer than those closer to the Galactic center.

Figure\,\ref{fig:SPZ_W147N12k_TM}(b) gives the evolution of the number
of stars in several of our models with $\Wo=4$ at various distances
 from the Galactic center and compares the results with a model in
which stellar evolution was not taken into account and in which stars
were not allowed to collide. All the calculations presented in this
figure were started with identical initial conditions. The only
difference between the various runs is the distance from the Galactic
center (varying from 34 to 150\,pc) and whether stellar evolution is
included in the calculation. In this figure, time is scaled with
respect to each model's initial relaxation time, i.e.:
\begin{equation}
	\trlx \propto N/8\log(0.4N).
\label{Eq:trlx2}\end{equation}

The various lines in Figure\,\ref{fig:SPZ_W147N12k_TM} do not overlap
perfectly, mainly due to stochastic differences in the moments when
binaries formed and interacted. The small deviations from the model in
which stellar evolution was not taken into account indicate that
stellar evolution in these models in unimportant.  If stellar
evolution were important, clusters at greater distances from the
Galactic center should dissolve more rapidly.  However, we see the
opposite trend: clusters farther from the Galactic center tend to
dissolve slightly more slowly; the model without stellar evolution
dissolves fastest. We have no ready explanation for this trend. Note
that for the models with stellar evolution this trend may be
attributed in part to collisions, as they tend to heat the cluster; so
clusters with higher collision rates (closer to the Galactic center)
would dissolve more quickly in terms of their initial relaxation time.
Collisional heating also causes the cluster to become less compact,
again reducing the cluster lifetime.

Alternatively the trend of longer lifetime at greater Galactocentric
distance may also be explainable in terms of the interplay between
stellar evolution, binary formation and stellar collisions.  These
possibilities require more detailed study, which is beyond the scope
of this paper; however, work is in progress to address this point. We
emphasize that, although the trend looks like a subtlety, it may play
an important role in the dynamical evolution of compact star clusters.

\begin{figure}[htbp!]
\hspace*{1.cm}
\psfig{figure=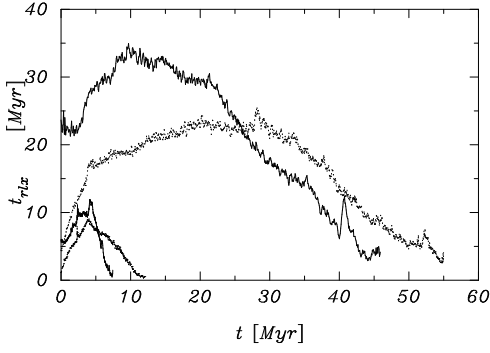,width=7.5cm,angle=-90}
\caption[]{Evolution of the half-mass relaxation time for clusters
lying 34\,pc and 150\,pc from the Galactic center, and with \Wo=1
(solids) and \Wo=7 (dotted lines; the \Wo=7 models live linger than
the \Wo=1 models). The models with \Wo=4 and those at $R_{\rm rg}
=90$\,pc are not given.}
\label{fig:SPZ_W147R34N12k_Ttrlx}
\end{figure}

Figure\,\ref{fig:SPZ_W147R34N12k_Ttrlx} shows the time variation of
the relaxation time for the models at 34\,pc and 90\,pc from the
Galactic center.  Note that the relaxation time measured some time
after zero age bears little information about the clusters' initial
conditions.  The behavior of the half-mass relaxation time is
qualitatively similar to that found in less compact clusters at
greater distances (6--12\,kpc) from the Galactic center (Portegies
Zwart et al.\, 2001b)---it first rises by a factor of a few as the
cluster expands, then slowly decreases as the cluster loses mass.

\begin{figure}[htbp!]
(a) \psfig{figure=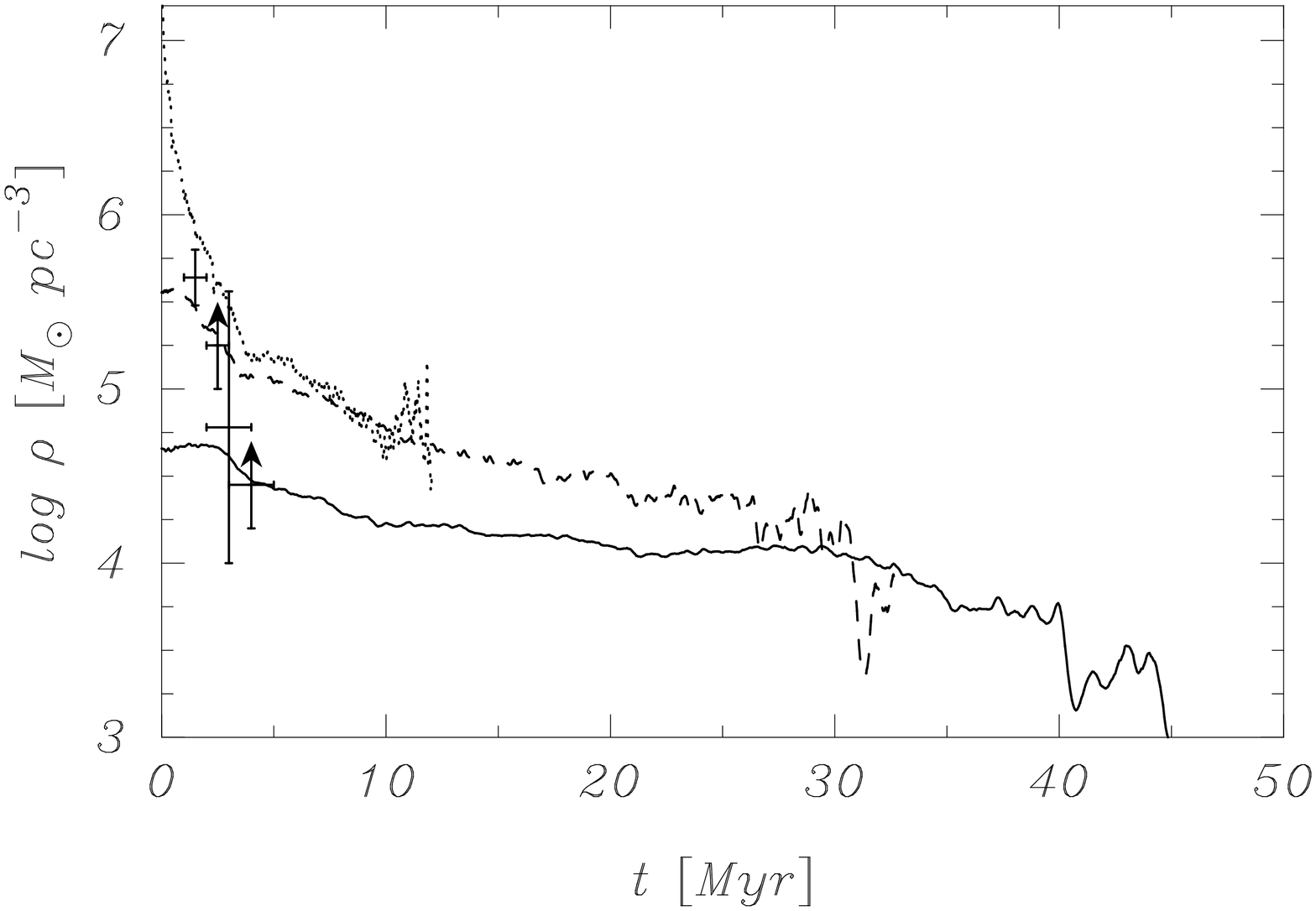,width=7.5cm,angle=-90}
(b) \psfig{figure=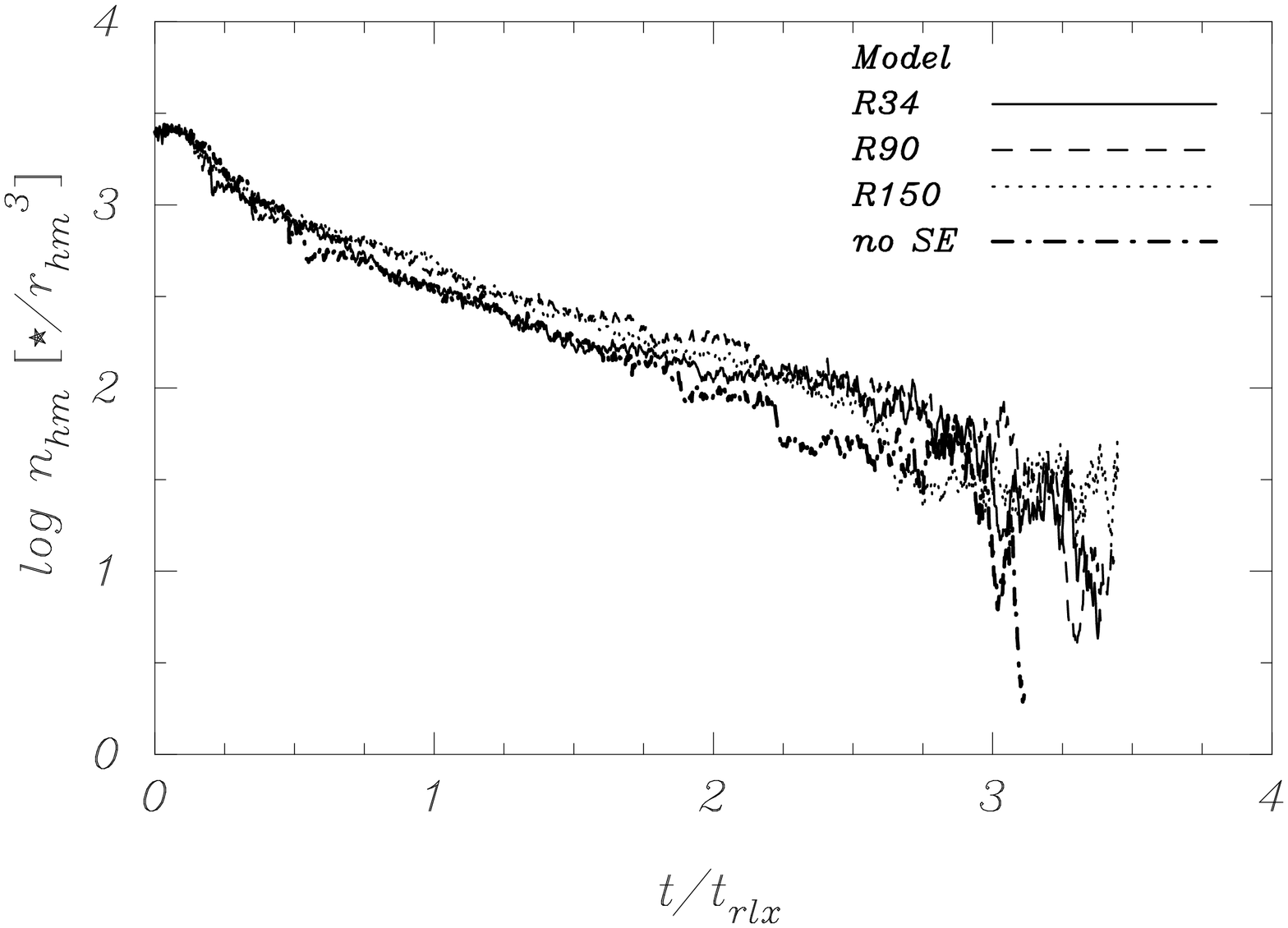,width=7.5cm,angle=-90}
\caption[]{Evolution of the half mass density for clusters.  (a) panel
in astronomical units ($\msun {\rm pc^{-3}}$ for the models R150W1
(solid), R90W4 (dashes) and R34W7 (dots).  Error bars give the values
for age and core density listed in Table\,\ref{Tab:observed}.  (b) the
density evolution of the models with \Wo=4 but now time is in units of
the half-mass relaxation time and density in units of star per half
mass radius cubed [see also panel (b) in
Fig.\,\ref{fig:SPZ_W147N12k_TM}].  }
\label{fig:SPZ_W147R34N12k_TRL}
\end{figure}

Figure\,\ref{fig:SPZ_W147R34N12k_TRL}(a) shows the evolution of the
mean densities within the half mass radius in models R150W1, R90W4 and
R34W7.  The initial half mass densities for these models were very
different, ranging about three orders of magnitude.  At later times
the differences in the half mass densities become much smaller; after
five million years the difference has decreased by about a factor of 5.

As in Fig.\,\ref{fig:SPZ_W147N12k_TM},
Figure\,\ref{fig:SPZ_W147R34N12k_TRL}(b) also shows the densities of
the models as functions of time scaled to the initial relaxation time
(see Portegies Zwart et al 2001a and Baumgardt
2001).\nocite{2000astro.ph.12330B} The trend visible in
Fig.\,\ref{fig:SPZ_W147N12k_TM} is somewhat obscured due to random
fluctuations in the density. 

\begin{figure}[htbp!]
\psfig{figure=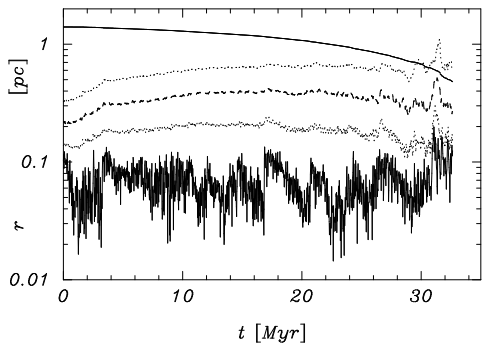,width=7.5cm,angle=-90}
\caption[]{Evolution of the core radius (lower solid) and the 25\%,
50\% and 75\% Lagrangian radii (dots, dashes and upper dotted lines,
respectively) for model R90W4.  The top line shows the
instantaneous Jacobi radius of the cluster.}
\label{fig:SPZ_W4R34N12k_TRc}
\end{figure}

Figure\,\ref{fig:SPZ_W4R34N12k_TRc} shows the evolution of the core
radius, selected Lagrangian radii, and the Jacobi radius \rLf, for
model R90W4.  Core collapse occurs within $\sim 1.2$\,Myr followed by
a gradual overall expansion of the cluster.  Similar behavior is
evident in the isolated clusters considered by Portegies Zwart et
al.~(1999).\nocite{pzmmh99} The post-collapse expansion stops after a
few million years, by which time the relaxation time has reached its
maximum value (see Fig.\,\ref{fig:SPZ_W147R34N12k_Ttrlx}).
Subsequently, the Lagrangian radii decrease as the Jacobi radius
shrinks and the cluster dissolves.

\subsection{Evolution of the binding energy}

\begin{figure}[htbp!]
(a) \psfig{figure=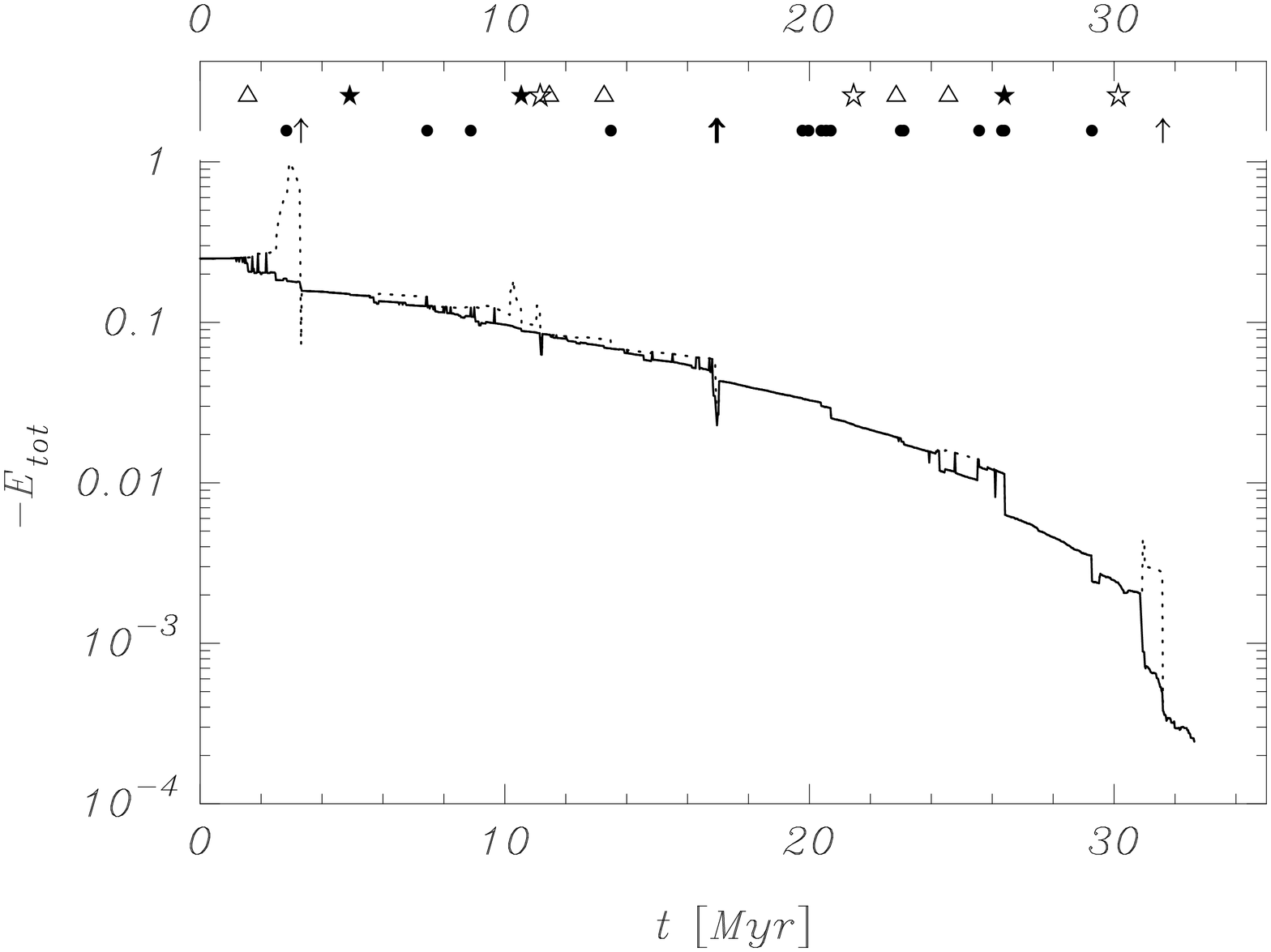,width=7.5cm,angle=-90}
(b) \psfig{figure=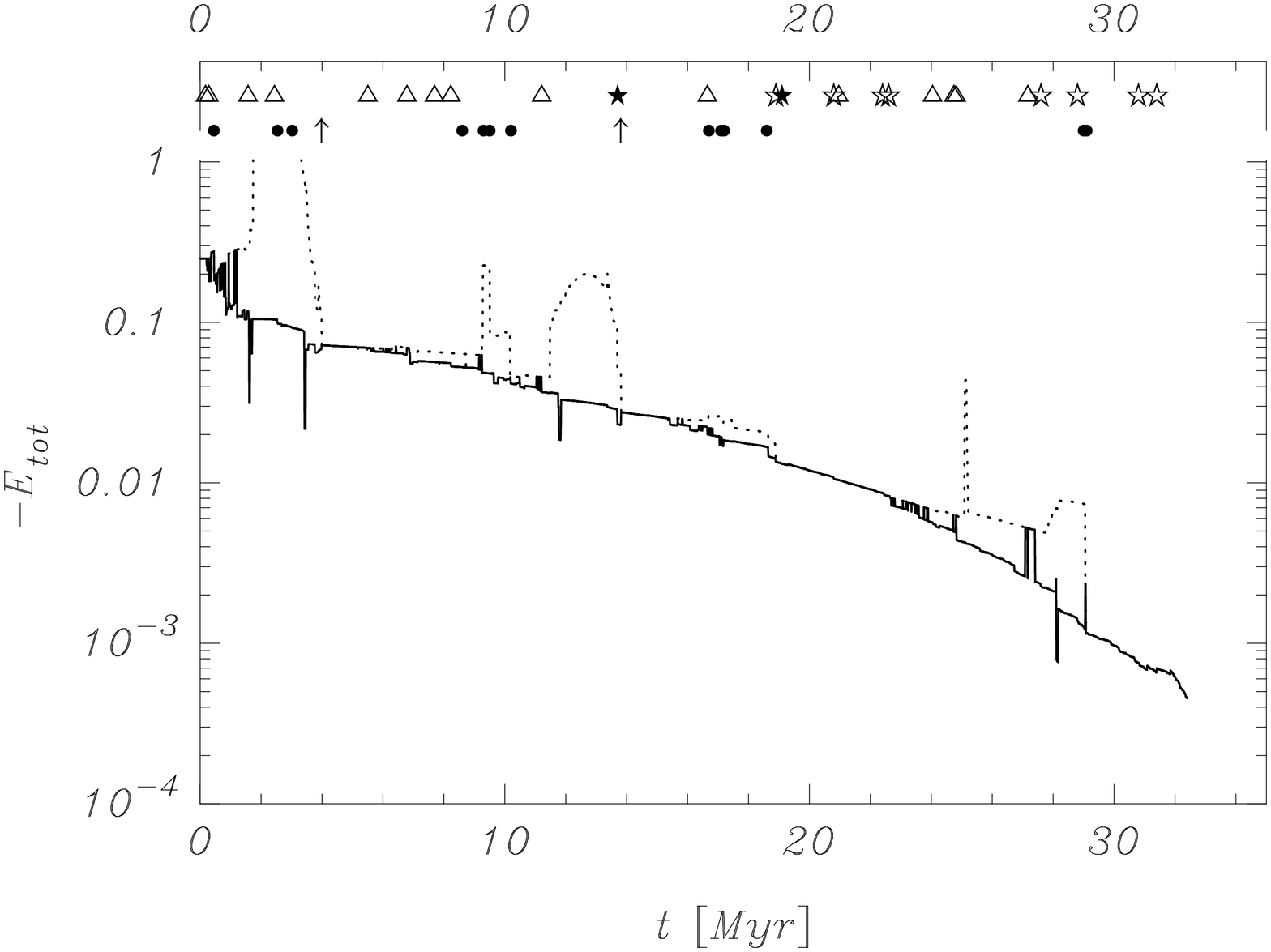,width=7.5cm,angle=-90}
\caption[]{Evolution of the binding energy for models R90W4 (a) and
R90W7 (b).  Solid lines include the binding energies of all stars
(only the center of mass energy of binaries and higher order systems
are included) the dotted lines include the binding energies of the
binaries. Initially $E_{\rm tot} = -0.25$ (see
Sec.\,\ref{Sect:Units}).  The two lines with symbols in the
horizontal bar above the figure indicate important happenings in the
cluster lifetime.  Triangles indicate the moment at which a binary or
triple is formed by a three- (or more) body interaction, the filled
and open asterisks indicate the formation of a black hole and neutron
star, respectively.  The filled circles indicate the moment a
collision occurs and arrows pointing upward indicate the moment then a
hard binary is ejected from the cluster.  All these events are
reflected in the evolution of the binding energy.  }
\label{fig:SPZW4R34N12k_Ebin}
\end{figure}

Figure\,\ref{fig:SPZW4R34N12k_Ebin} shows the evolution of the binding
energy ($-E_{\rm tot}$, in scaled {\nbody} units) for models R90W4 and
R90W7, which are selected for their similar lifetimes.  (Some
deviations between the solid and the dotted lines are a result of our
definition of a binary. Here we consider a binary to be a pair of
bound stars for which the internal forces are a hundred times higher
than those due to the nearest neighbor.)  Initially, each cluster has
$E_{\rm tot} = -0.25$ (see Sec.\,\ref{Sect:Units}), but due to
escapers, binary activity, stellar mass loss, mergers, supernovae,
etc., the total energy changes.  A cluster ceases to exist when its
binding energy becomes positive.  The two clusters exhibit quite
similar global evolution.  Model R90W4 shows an initial ``plateau'' of
a few Myr during which mass segregation occurs, driving the cluster
into the core collapse phase (see Fig.\,\ref{fig:SPZ_W4R34N12k_TRc}).
For model R90W7 this phase is hardly noticeable as the cluster
experiences several strong encounters and a few collisions early in
its evolution.

During the first deep core collapse the binding energy fluctuates
rapidly because binaries frequently harden and single stars are
ejected following three-body encounters.  The first collisions happen
during this phase (see $\bullet$ in Fig.\,\ref{fig:SPZW4R34N12k_Ebin})
and in both models mass transfer occurs in a dynamically formed
binary, giving rise to an excursion in the binding energy (see the
dotted line in Fig.\,\ref{fig:SPZW4R34N12k_Ebin}). Core collapse is
followed by the ejection of a hard binary (arrows in
Fig.\,\ref{fig:SPZW4R34N12k_Ebin}).

After the first collapse of the core the binding energies of both
clusters decrease steadily with some strong fluctuations caused by
close two- and three-body encounters (triangles in
Fig.\,\ref{fig:SPZW4R34N12k_Ebin}), high-velocity escapers (arrows),
stellar collisions ($\bullet$), mass transfer in close binaries and
supernovae ($\star$).

Model R90W7 experiences much more binary activity than model R90W4.
This can be seen from the many departures of the dotted line in
Fig.\,\ref{fig:SPZW4R34N12k_Ebin} from the solid curve.  Such
excursions are the result of binary activity, such as mass transfer.
A stable phase of mass transfer generally results in an increase in
the binding energy of the binary. Such binaries first become very hard
as the orbital period decreases, then soften when the mass of the
donor drops below that of the accreting star.  One example of this is
clearly visible in panel (b) of Fig.\,\ref{fig:SPZW4R34N12k_Ebin},
around $t=11$\,Myr. The phase of mass transfer stops after about
3\,Myr, indicating that it is a case A (Kippenhahn \& Weigert
1967),\nocite{kw67} i.e.~stable on a nuclear time scale (see also
Portegies Zwart et al.\, 2001b).\nocite{2001MNRAS.321..199P} This
episode stops when the donor leaves the main sequence, becomes an
envelope-depleted helium burning star and finally explodes in a
supernova (filled star). The mass loss in the supernova, forming a
black hole, causes the binary to be ejected from the cluster (up
pointed arrow).

The spikes to lower binding energy in both models R90W4 and R90W7 are
the result of stars which receive a high velocity following a strong
encounter. Such a high-energy encounter may decrease the binding
energy of the entire cluster for as long as the star is considered a
cluster member (within 3{\rLf} from the cluster center).  Model R90W7
produces many more high-velocity escapers than model R90W4, consistent
with the greater dynamical activity within the more compact cluster.

\subsection{Evolution of the total luminosity}\label{sect:Ltot}

The total luminosities of the model clusters decrease with time as the
clusters dissolve in the Galactic tidal field and the most massive
stars evolve off the main sequence and become dark remnants.

\begin{figure}[htbp!]
a) \psfig{figure=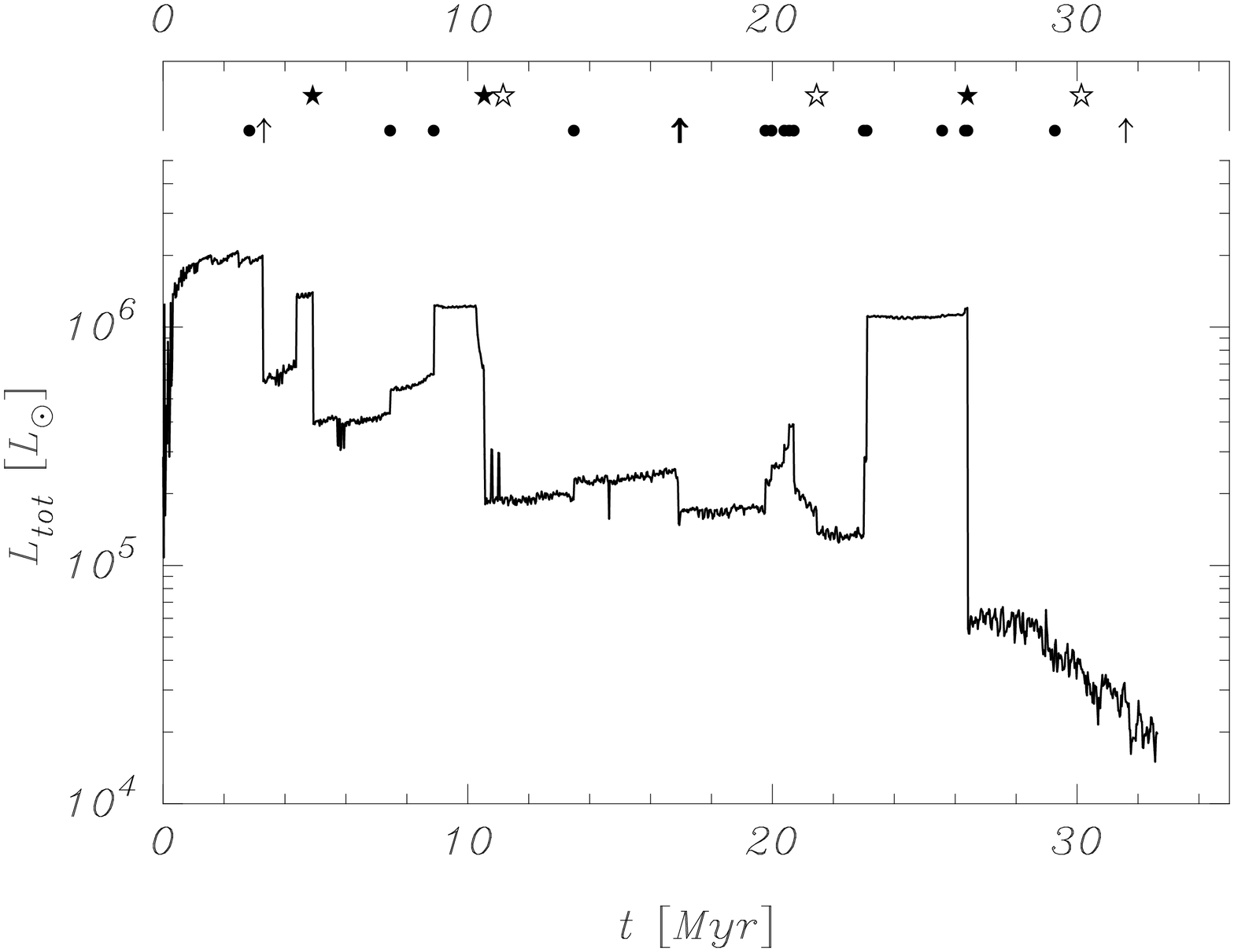,width=7.5cm,angle=-90}
b) \psfig{figure=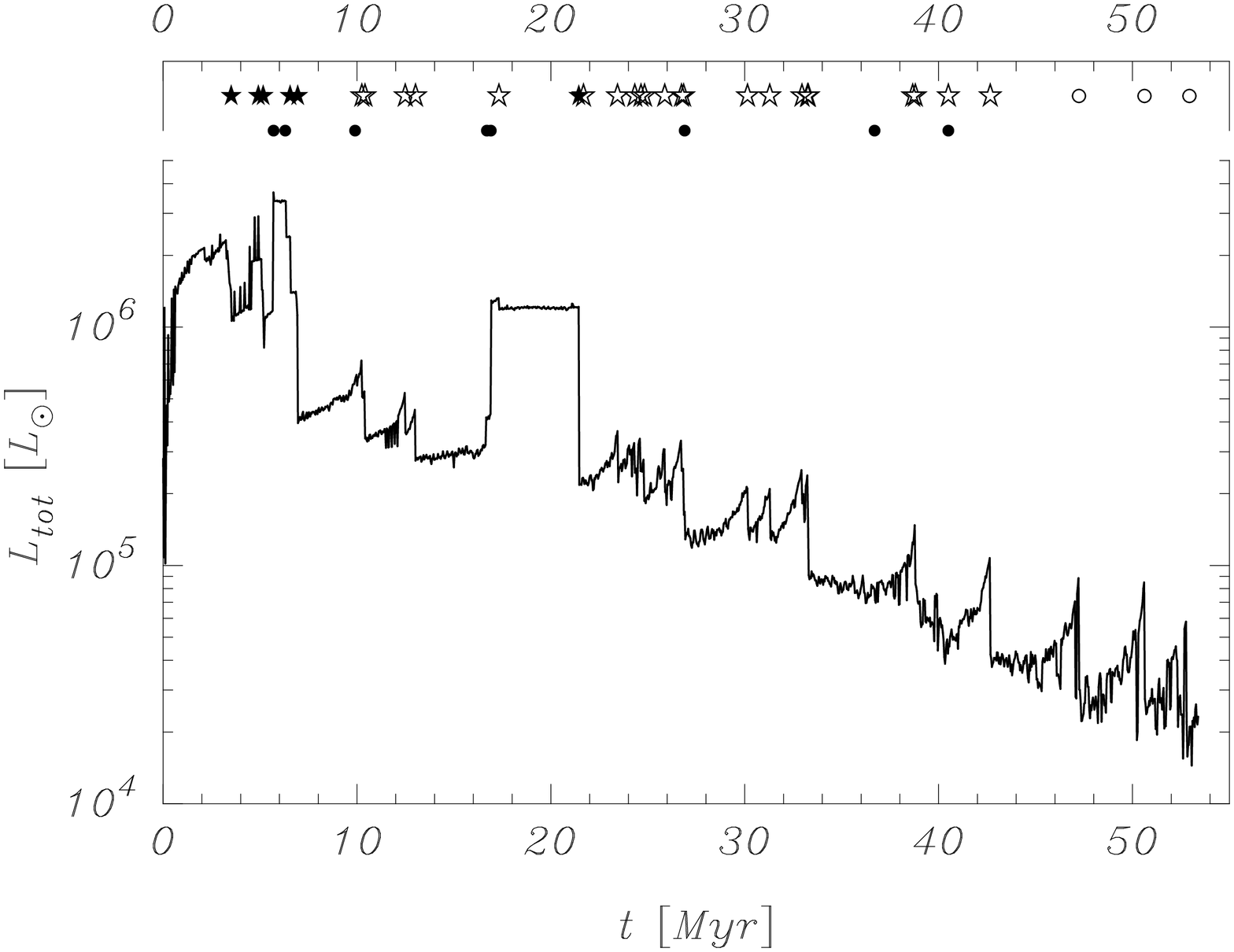,width=7.5cm,angle=-90}
\caption[]{Evolution of the visual luminosity of all stars within
3\rLf\, for the model R90W4 (a) and model R150W4 (b).  Both models
were started with the same random seed and therefore have identical
stellar masses, positions and velocities [in scaled \nbody\, units].
Solid line include the luminosity of all stars.  The two lines with
symbols in the horizontal bar above the figure indicate important
happenings in the cluster lifetime, Fig.\,\ref{fig:SPZW4R34N12k_Ebin}
gives the explanation of the symbols (we omitted the information about
close encounters since they bear little information [in panel b we
also omitted the escapers]). The open circles in the upper bar of
panel (b) indicate the moment when a super giant forms a white dwarf.
}
\label{fig:TL_W4150}
\end{figure}

Figure\,\ref{fig:TL_W4150} presents the integrated visual luminosity
for the models R90W4 (see also Fig.\,\ref{fig:SPZW4R34N12k_Ebin}a) and
R150W4 (both models had identical initial stellar masses [in \msun],
positions and velocities [in scaled \nbody\, units]).  Initially the
luminosity greatly exceeds $10^6$\,\lsun, then drops steadily by more
than two orders of magnitude as the cluster ages. After about
4\,Myr the luminosity suddenly drops due to the explosion of the most
massive stars (for model R150W4) or because a few massive stars escape
 from the cluster as the binary in which they reside is ejected (for
model R90W4, see also Fig.\,\ref{fig:SPZW4R34N12k_Ebin}).

Occasionally the luminosity again exceeds $10^6$\,\msun\, as a runaway
collision product leaves the main sequence and becomes a Luminous Blue
Variable or a Wolf-Rayet star. (\PaperIII\, discusses the collision
rate in models with comparable initial conditions. A summary is given
in Sec.\,\ref{sect:comparison}.)  In model R90W4
(Fig.\,\ref{fig:TL_W4150}a) this happens around 9.5\,Myr and
23--25\,Myr and for model R150W4 (Fig.\,\ref{fig:TL_W4150}b) near
$\sim 6$\,Myr and 16--22\,Myr. In these episodes the total luminosity
of the entire star cluster is dominated by a single star.  This
episode suddenly stops when the star explodes or escapes from the
cluster.  The regular rises and sudden drops in luminosity (for
example near $t=33$\,Myr and $t=47$\,Myr in Fig.\,\ref{fig:TL_W4150}b)
are caused by single stars which ascend the giant branch and then
explode in supernovae (open and filled stars) or form a white dwarf
(circles).  The essential point, however, is that the cluster is
brightest at early times. Later increases in brightness may be caused
by the rejuvenation of stars in collisions, but in general the total
luminosity decreases steadily until the cluster dissolves completely.

\subsection{Fraction of massive stars in the cluster center}

\begin{figure}[htbp!]
\hspace*{1.cm}
\psfig{figure=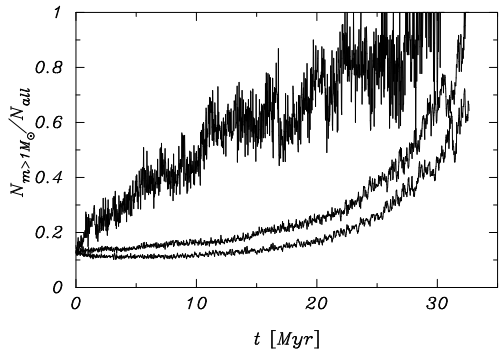,width=7.5cm,angle=-90}
\caption[]{Fraction of stars having masses $>1$\,{\msun} within the
0--5\% (upper line), 10--50\% (middle line), and 50--90\% (lower line)
Lagrangian radii, for model R90W4.  
}
\label{fig:SPZW4R90N12k_Tnla}
\end{figure}

Figure\,\ref{fig:SPZW4R90N12k_Tnla} shows the evolution of the
fraction of stars having masses greater than 1\,\msun, within the 5\%
Lagrangian radius and near the 50\% and 75\% Lagrangian radii of model
R90W4.  Mass segregation causes the central mass function to flatten
rapidly, but it takes considerably longer before the outer region of
the cluster is significantly affected.  After $\sim4$ million years,
the mass function within the inner 5\% Lagrangian radius contains
about 4 times as many stars with masses exceeding $1$\,{\msun}
(relative to the local number density) than does the rest of the
cluster.  The mass function near the half-mass radius remains
comparable to that of the cluster as a whole (see also Vesperini \&
Heggie 1997).\nocite{1997MNRAS.289..898V} The outer 75\% of the
cluster becomes slightly depleted of high-mass stars.  Within one
initial half-mass relaxation time, the fraction of stars having
$m>1\,\msun$ in this region falls by about a factor of 3.  At later
times, the fraction of high-mass stars throughout the cluster
increases due to the preferential escape of low-mass stars; at
disruption, the cluster is rich in high-mass stars, while low-mass
stars are depleted.  This is in agreement with the findings of
Takahashi \& Portegies Zwart (2000),\nocite{2000ApJ...535..759T} who
concluded that the observed flat mass function in the globular cluster
NGC\,6712 indicates that this cluster is close to dissolution.

\subsection{Has the Arches cluster an unusually flat mass
function?}\label{sect:mass_function}

Figer et al.~(1999b)\nocite{1999ApJ...514..202F} studied the mass
function of the Arches cluster in two projected annuli, the inner one
spanning 2.5 to 4.5 arc seconds from the cluster center (0.098 to 0.18
pc, assuming a distance of 8 kpc), and the outer one extending from
 from 4.5 to 7.5 arc seconds (0.18 to 0.29 pc).  They found that the
cluster mass function is much flatter than the Salpeter distribution
(a power law with exponent $x = -2.35$), and that the mass function in
the inner annulus is even flatter ($x = +1$ to $-1.5$) than that in
the outer parts ($x = -1.5$ to $-2.0$).  They estimate completeness
down to 20\,\msun\, in the inner annulus, and down to 10\,\msun\, in
the outer one.  The total number of stars in the inner annulus is 50
down to 20\,\msun, for a total mass of 2173\,\msun.  The total number
of stars in the outer annulus is 122 down to 10\,\msun\, resulting in
a total mass of 3164\,\msun.

\begin{figure}[htbp!]
a)\psfig{figure=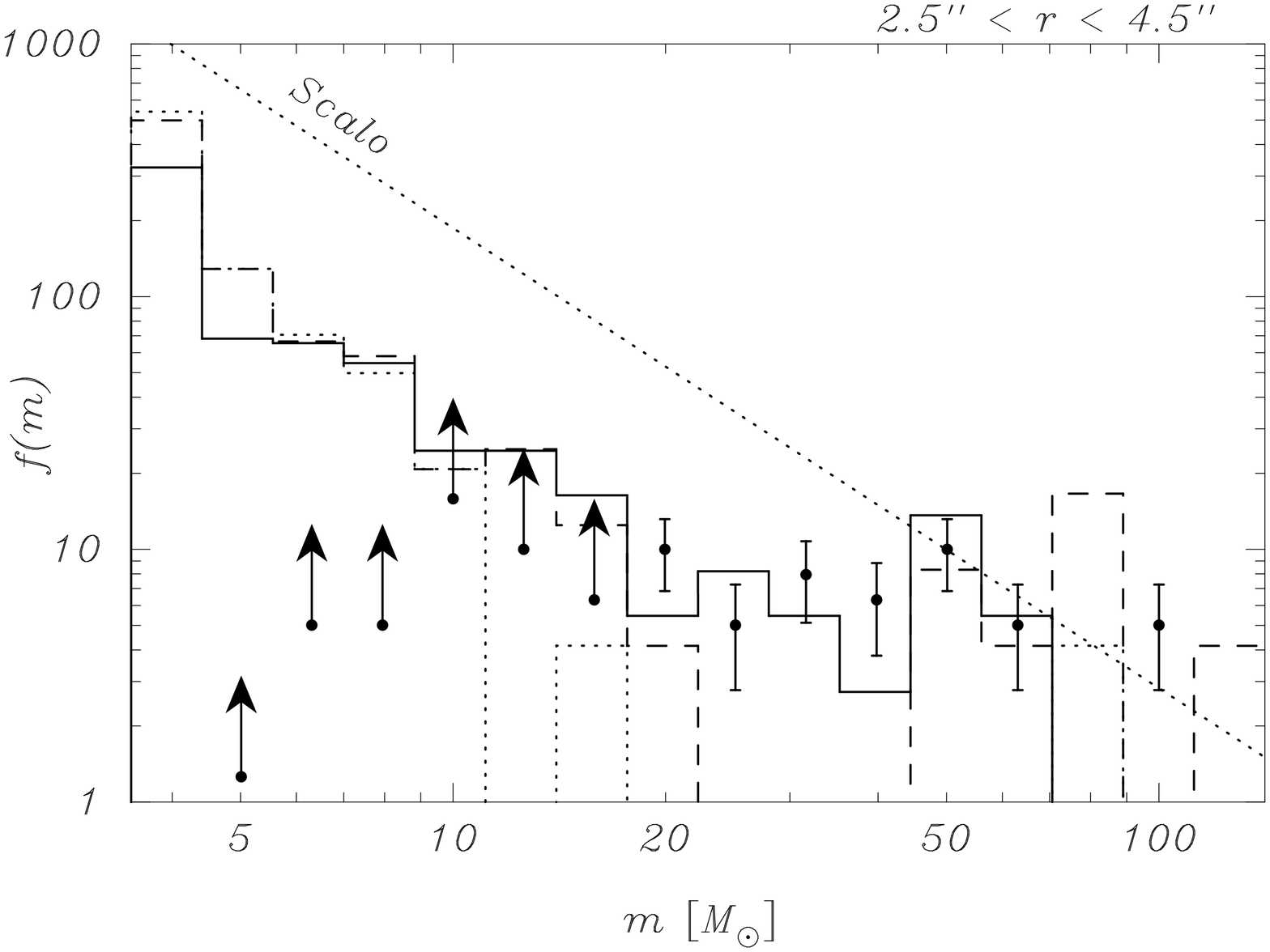,width=7.5cm,angle=-90}
b)\psfig{figure=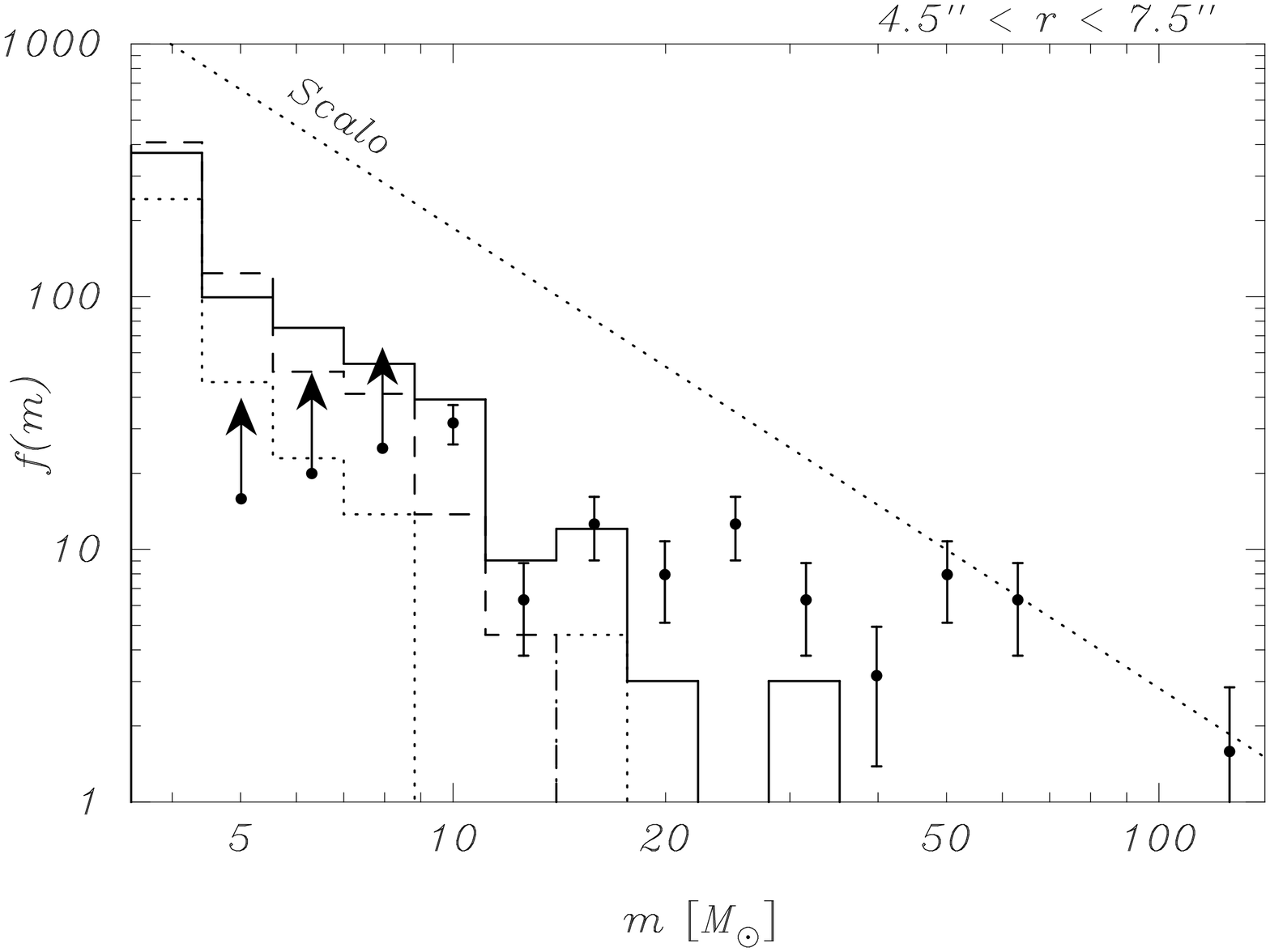,width=7.5cm,angle=-90}
\caption[]{ Mass function for models R150W4 (solid) and R90W4 (dashes)
and R34W4 (dots) at an age of 3\,Myear.  The initial mass function of
the whole cluster at birth is given by the dotted line (Scalo 1986).
a) the mass functions at the projected inner annulus ($2.5'' < r <
4.5''$).  b) the mass function in the outer annulus ($2.5'' < r <
4.5''$) from the cluster center.  These annuli correspond to those in
which Figer et al. (1999a) observed the mass function for this cluster
(bullets).  Arrows indicate lower limits due to incompleteness.  
}
\label{fig:MF_F205}
\end{figure}

Figure\,\ref{fig:MF_F205} shows the observed mass function in the
inner and outer annuli and compares them with the mass functions of
our models with $\Wo=4$ at 34\,pc (dots), 90\,pc (dashes) and 150\,pc
(solid line) from the Galactic center. The projected distance was kept
constant at 34\,pc.  The comparison in Fig.\,\ref{fig:MF_F205} is
performed at an age of 3\,Myr. Each calculation was repeated three
times to improve statistics. We further compensate for the uncertainty
in cluster ages and distances from the Galactic center by using annuli
which are somewhat narrower (from 0.098\,pc to 0.139\,pc and from
0.18\,pc to 0.235\,pc for the inner and outer annuli, respectively)
than the observed annuli (from 0.098\,pc to 0.18\,pc for the inner
annulus and from 0.18\,pc to 0.29\,pc for the outer annulus).  We
normalize to match a star cluster with $10^5$ stars\footnote{When
comparing two stellar systems with a different number of stars a
choice must be made about the time at which the comparison is carried
out. The time can be expressed in units of the initial crossing time
(or in million years as in Fig.\,\ref{fig:MF_F205}), or in terms of
the initial relaxation time of the cluster. For a purely dynamical
comparison the initial relaxation time may be more natural (for a
discussion see Portegies Zwart et al.\, 1998).}.

The mass function for the cluster farther from the Galactic center is
in better agreement with the observations; the models at $\rgc=34$\,pc
do not reproduce the observations well.  In addition, the inner
annulus is much better represented by both sets of models than is the
outer annulus, probably because the latter is much more sensitive to
the dynamical evolution of the cluster and the adopted distance from
the Galactic center. The top end of the mass function
($\apgt50$\,\msun) is underrepresented in both annuli.  This
discrepancy is caused in part by the steep slope at the upper end of
our selected initial mass function (which has an exponent -2.8, where
Sapleter is -2.35).  We simply have not enough massive stars in our
simulations.  For stars more massive than $\apgt 50$\,\msun, the
initial mass function in young star clusters may well be flatter than
than adopted by Scalo (1986).

This method of comparing mass functions is very sensitive to the total
mass of the cluster, its age, and the actual distance between the
cluster and the Galactic center. These three parameters are coupled
via the relaxation time, and small changes in any of them may have a
profound effect on the derived mass function for the real cluster.

In order to compare our model calculations more directly with the
observations we introduce the ratio of the mean mass in the outer
annulus to that in the inner annulus, which for the observations is
$f_{o/i} \equiv \langle m \rangle_{\rm out}/ \langle m \rangle_{\rm
in} \simeq 25.9\msun/43.5\msun = 0.597$.  By presenting this value as
a function of time (measured in units of the initial relaxation time
of the cluster), we remove much of the observational bias.

This mean mass ratio $f_{o/i}$ is a characteristic of the cluster.
For a Scalo (1986)\nocite{scalo86} mass function at zero age and in a
system with a homogeneous density the value of this fraction is
$f_{o/i} = 0.501$\footnote{The initial value of $f_{o/i}<1$ because of
the different lower cutoffs in the nominator and the denominator.  The
selected cutoffs are identical to those in the observations, which
are more affected by crowding in the inner annulus.}.  As more massive
stars sink to the cluster center due to mass segregation the value of
$f_{o/i}$ changes. The value of $f_{o/i}$ also changes due to the
variation of the density profile in the cluster, and therefore is also
a function of the distances from the cluster center at which $f_{o/i}$
is measured.  
The shapes of these curves, however, are rather insensitive to the initial
density profile because the evolution of the cluster mass
(Fig.\,\ref{fig:SPZ_W147N12k_TM}b) and density
(Fig.\,\ref{fig:SPZ_W147R34N12k_TRL}b) do not depend sensitively on
the adopted density profile.
Because of the uncertainties in the masses and hence the
length scales of our models, we compute $f_{o/i}$ at various distances
from the centers of our model ($\Wo=4$) clusters as functions of time.

\begin{figure}[htbp!]
\hspace*{1.cm}
\psfig{figure=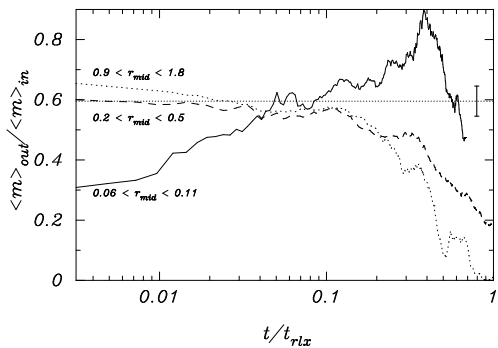,width=7.5cm,angle=-90}
\caption[]{The values for $f_{o/i}$(the mass ratio of the mean mass in
the outer annulus divided by the mean mass in the inner annulus) as a
function of the time in units of the initial half-mass relaxation
time.  Each of the various lines combines the results for two of the
measured values for $f_{o/i}$. The solid line gives the sum of inner
two values (with $r_{\rm mid}=0.06$ and 0.11), dashed line for the
intermediate range ($r_{\rm mid}=0.2$ and 0.5) and the dotted line
gives the value of $f_{\rm o/i}$ for the outer annuli ($r_{\rm
mid}=0.9$ and 1.8).  The horizontal dotted line gives the measured
value of $f_{\rm o/i}$ as measured by Figer et al.\, (1999a) with the
Poissonian error bar (at right).  }
\label{fig:MoMi_vs_Trlx}
\end{figure}

We selected a total of 6 sets of inner and outer annuli, starting
close to the cluster center and moving progressively outward. For the
innermost annulus we selected $r_{\rm in} = 0.003125$, $r_{\rm mid} =
0.05625$ and $r_{\rm out} = 0.09375$ (in units of the initial virial
radius) and we increased the radii by factors of two for subsequent
annuli out to $r_{\rm in} = 1.0$, $r_{\rm mid} = 1.8$ and $r_{\rm out}
= 3.0$. For each of our \nbody\, calculations we computed $f_{o/i}$
for all combinations of 6 inner and outer annuli. To improve
statistics we then combined the values for the inner two (solid line
in Fig.\,\ref{fig:MoMi_vs_Trlx}), the middle two (dashes) and the
outer two (dotted line) annuli.

Figure\,\ref{fig:MoMi_vs_Trlx} plots these values of $f_{o/i}$ as a
functions of time for our model calculations with $\Wo=4$ and compares
them with the observed mass function. We combined all models with
$\Wo=4$ to improve the statistics at the top end of the mass function.
We computed a total of 14 models with these parameters, of which 8
were run for only 0.91 initial relaxation times and 6 continued until
they dissolved (more than 3\,\trxh).
The various models were scaled to the initial relaxation time before
being superimposed, as explained in Sec.\,\ref{sect:results}. At later
times ($t \apgt \trxh$) some effect from stellar evolution may be
influencing this scaling, but since we are mostly concerned here with
earlier times this does not affect our conclusions.  The vertical bar
representing the observations may be shifted along the dotted
line. The best fit for all annuli is obtained in the interval $0.05
\aplt \trxh \aplt 0.2$.

The value of $f_{o/i}$ for the inner annulus in
Fig.\,\ref{fig:MoMi_vs_Trlx} is much lower than for the outer
annuli, but $f_{o/i}$ steadily increases to match the outer annuli near
$t \simeq 0.05$\,\trxh. In the same time interval $f_{o/i}$ for the
outer annuli decreases somewhat. This effect is caused by mass
segregation, as the most massive stars sink to the cluster center more
rapidly than somewhat less massive stars. On their way to the cluster
center the most massive stars pass through the various annuli. The
loss of a few massive stars has little effect on the value of
$f_{o/i}$ in the outer part of the cluster as these annuli contains
many more stars than the inner annuli. The arrival of a few massive
stars in the cluster center, however, carries a relatively larger
contribution to the total mass in high mass stars to the inner
annulus. The inner $f_{o/i}$ rises therefore more quickly than the
value in the outer annuli decreases.

At later time $t \apgt 0.2\,\trxh$ the value of $f_{o/i}$ rises sharply
for the inner annulus while for the outer annuli $f_{o/i}$ continues
to decrease. The reason for this is the continuing effect of mass
segregation; the more massive stars pile up in the cluster center
while the clusters' outer parts become depleted of massive stars.
After $t \simeq 0.4\,\trxh$, $f_{o/i}$ for the inner annuli (solid
line in Fig.\,\ref{fig:MoMi_vs_Trlx}) starts to decrease rapidly.  The
sharp peak in this curve is the result of the two competing effects:
mass segregation causing $f_{o/i}$ to increase, and stellar evolution
resulting in its decline.  This strong decline in $f_{o/i}$ is caused
by the onset of stellar evolution as the model with the longest
relaxation time ($\trxh = 13$\,Myr at $\rgc = 150$\,pc, see
Tab.\,\ref{Tab:N12kinit}) starts to run out of massive stars. The
first decline starts at $t \simeq 0.4\,\trxh$, corresponding to an age
of 5.2\,Myr for the $\trxh = 13$\,Myr model. This time corresponds to
the nuclear lifetime of a 34\,\msun\, star.  The lower mass limit used
for the outer annuli is 20\,\msun. Such a star has a nuclear burning
lifetime of about 9.1\,Myr i.e.: by the time $t\apgt 0.70$\,\trxh\,
all stars more massive than 20\,\msun\, have experienced a supernova
and $f_{o/i}$ has consequently diminished.  (The slightly later time
at which the outer annuli run out of stars is caused by the
inhomogeneous mixture of models used to make
Fig.\,\ref{fig:MoMi_vs_Trlx}.)


The observed value of $f_{o/i}$ (vertical bar) may be moved along the
horizontal dotted line until it matches any of the curves representing
the model calculations. A matching result indicates that the observed
mass function at these annuli agrees with a cluster at the appropriate
age in units of the initial relaxation time.  We illustrate this with
a few examples.

At later times ($t\apgt 0.1\,\trxh$) the observations fit quite well
with the solid line in Fig.\,\ref{fig:MoMi_vs_Trlx}, representing the
innermost annuli. Initially, high-mass stars are under sampled in the
inner part of the cluster, and only at later times does mass
segregation cause them to sink to center of the potential well of the
system, explaining the initial rise in $f_{o/i}$ for the innermost
annuli.  If indeed the observed value of $f_{o/i}$ coincides with the
model calculations at the innermost part of the cluster, the observed
cluster must be older than about $t \apgt 0.1\trxh$. At an observed
age of the Arches cluster of 1--2\,Myr the initial relaxation time
must then have been $\trxh \aplt 20$\,Myr. Using the intermediate
annulus of $r_{\rm mid} = 0.18$\,pc to set the size scale of the
models, the half-mass radius of the Arches cluster is then $\rhm \apgt
1.6$---$3.0$\,pc.  A cluster with a half-mass radius of $\rhm \apgt
1.6$\,pc and a relaxation time $\trxh \aplt 20$\,Myr contains less
than a hundred stars.  Since the number of stars observed easily
exceeds this number we can firmly reject this solution: the observed
radius of $r_{\rm mid} = 0.18$\,pc then corresponds to $\aplt
2$\,\rvir.

Alternatively, we may observe the cluster when it is younger in terms
of its relaxation time.  In the very early evolution of the cluster
$t\aplt 0.05\trxh$ the observations are quite consistent with $0.2
\aplt r_{\rm mid} \aplt 1.8$ (the dashed as well as with the dotted
line in Fig.\,\ref{fig:MoMi_vs_Trlx}), but is inconsistent with
$r_{\rm mid} \aplt 0.11$.  The intermediate radius in the observations
corresponds to $r = 4.5''$ (or about $0.18$\,pc at a distance of
8\,kpc). With these values the cluster would be described accurately
by $\rvir = 0.1$--0.9\,pc at an age $\aplt 0.05\,\trxh$.  With an age
of the Arches cluster of 1--2\,Myr (see Tab.\,\ref{Tab:observed}) the
initial relaxation time of the cluster must be larger than 20--40
\,Myr.  With a half-mass radius $\rvir \aplt 0.9$ (corresponding to
$\rhm \simeq 0.75$\,pc), the cluster then has a mass of about
4\,000\,\msun\, in order to produce a half-mass relaxation time of
$\trxh \aplt 40$\,Myr. If $\rvir = 0.1$\,pc, the total mass is
$2.4\times 10^6$\,\msun\, and the relaxation time is 20\,Myr.

Taking the observed half mass radius of $\rhm \sim 0.2$\,pc and
correcting it for the effects of mass segregation and expansion during
its early dynamical evolution to $\rhm \sim 0.35$\,pc, a relaxation
time of 30\,Myr implies a cluster of about 64k stars and a total mass
of about 40\,000\,\msun.  These numbers would also agree with the
observed number of massive stars (50 stars with $m>10$\,\msun\, in the
inner annulus) if a normal Scalo (1986) mass function were adopted.
We thus see no reason to invoke a flatter than usual mass function to
explain the observations.  This is consistent with the suggestion of
Serabyn, Shupe \& Figer (1998)\nocite{1998Natur.394..448S} that the
Arches contains $\sim 10^5$ stars, far more than the 12k adopted in
our models or as suggested by Figer et al. (1999a), as we discuss in
Sec.\,\ref{Sect:Discussion}.

\begin{figure}[htbp!]
\hspace*{1.cm}
\psfig{figure=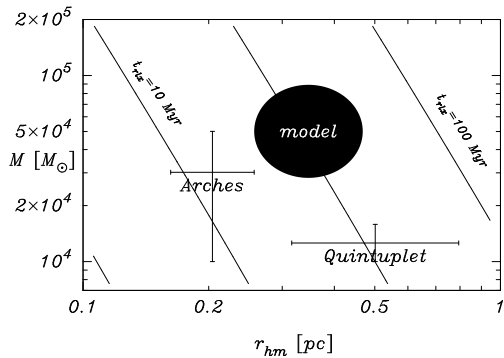,width=7.5cm,angle=-90}
\caption[]{Graphical representation of the constraints for our model
of the Arches cluster.  The horizontal and vertical axis give the half
mass radius and initial cluster mass, respectively. The diagonal lines
are of constant relaxation time at the half mass radius
(Eq.\,\ref{Eq:trlx}).  The two error crosses locate the observed
parameters for the Arches and Quintuplet star clusters.  The ellipse
identifies the parameter space in which our model calculations are
consistent with the observed mass function, number of high mass stars
and density structure of the observations for the Arches cluster (see
discussion in Sec.\,\ref{sect:mass_function}).  }
\label{fig:constraints}
\end{figure}

The limits on the initial parameters for the Arches cluster are
summarized in Fig.\,\ref{fig:constraints}. The shaded ellipse shows
the likely initial parameters for this star clusters.

Assuming that the density profile of the cluster is consistent with a
King model with $\Wo=4$--6 and adopting $\rhm = 0.35$, we derive a
Jacobi radius of 1.6\,pc--2.5\,\pc.  Consequently the distance from
the Arches cluster to the Galactic center is between 43\,pc and
91\,pc, for models with \Wo=4 and \Wo=6, respectively.  We conclude
that the Arches cluster is likely to be at a distance from the
Galactic center somewhat, but not much, greater than its projected
distance.  Table\,\ref{Tab:Arches} reviews our conclusions about the
Arches system.

\begin{table}[htbp!]
\caption[]{Derived parameters for the Arches cluster resulting from
our best values from our model calculations.  Assuming a normal mass
function comparable to that of the Solar neighborhood.  These numbers
are derived in Sec.\,\ref{sect:mass_function} and discussed in the
sections\,\ref{sect:density_profile} and \ref{sect:Rgc}.  }
\begin{flushleft}
\begin{tabular}{lr} \hline
parameter & value \\ 
\hline
$M$           & $\sim 40\,000$\,\msun \\ 
$N$           & $\sim 65\,000$\,stars  \\
\rgc	      & 43 -- 91\, pc       \\
\trxh	      & 20 -- 40\, Myr       \\ 
age	      & $0.05\pm 0.02$\,\trxh \\
\rhm	      & $0.35 \pm0.05$\, pc     \\
\rvir	      & $0.42 \pm0.05$\, pc     \\
\Wo	      & 4 -- 6              \\ 
\rtide	      & 1.6 -- 2.5\, pc      \\ \hline
\hline
\end{tabular}
\end{flushleft}
\label{Tab:Arches} \end{table}

\subsection{The density profile}\label{sect:density_profile}

\begin{figure}[htbp!]
\hspace*{1.cm}
\psfig{figure=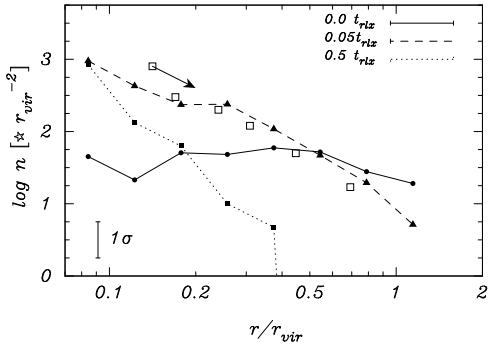,width=7.5cm,angle=-90}
\caption[]{Projected density profile of the Arches cluster for stars
with $m>20$\,\msun\, from the observations of Figer et al.\, (1999a;
data from Kim et al.\, 2000---open squares).  These
data were scaled to a virial radius of 1\,pc. The arrow
starting from the upper left most square indicates the direction in
which to shift these symbols when the size scaling (in arc seconds) is
increased by $1\sigma$.  They adopted the inner annulus between $2.5 <
r < 4.5$ arc seconds ($0.098 < r < 0.18$\,pc) and an outer annulus
between $2.5 < r < 4.5$ arc-seconds ($0.18 < r < 0.29$\,pc).
The various other lines and symbols give the results of our \nbody\,
calculations with $\Wo=4$. The various lines represent the density
profile at the moment indicated (top right corner).  These moments are
presented in units of the initial relaxation time of our models,
ranging from zero age (solid line) to 0.5\,\trxh.  
The densities for our model calculations were increased with a factor
64/12 to account for the larger number of stars expected in these
clusters. 
In the lower left corner is a $1\sigma$ Poissonian error bar for
reference.  
}
\label{fig:rvir_rhosq}
\end{figure}

Figure\,\ref{fig:rvir_rhosq} shows the projected density profile of
our model clusters with \Wo=4 for stars having masses $m\apgt 20$\,\msun\,
at various instants.  The data are presented in virial units
(horizontal axis in units of \rvir, and the vertical axis in units of
stars per $r^2_{\rm vir}$).  The various curves (and symbols) represent
the density distribution of the model clusters at zero age (solid
line), at $t=0.05$ initial relaxation times (dashes), and at
$t=0.5$\trxh (dotted line).  Mass segregation causes the density
profile of the older clusters to steepen with time, tidal
stripping causes the global density to decrease.

The open squares in Fig.\,\ref{fig:rvir_rhosq} present the observed
density distribution for the Arches cluster (Figer et al.\, 1999a).
These data are given in arc seconds (or parsec when scaled to the
appropriate distance). We plot them on the same figure assuming that
the observed median radius of 0.18\,pc (4.5 arc seconds) corresponds to
0.18\rvir, i.e.: that the virial radius of the Arches cluster is
1\,pc. This may be an overestimate, as discussed in the previous
section. There is some freedom in shifting the observed points along
the arrow, which then changes the scaling for the cluster.

The observed points fit best with the dashed line and are not
consistent with the solid (zero age) or the dotted ($t=0.5\trxh$)
lines.  The projected density profile of the observed Arches cluster
thus suggests an age of about $t\simeq 0.05\,\trxh$, in agreement with
our findings in Sec.\,\ref{sect:mass_function}.

\subsection{Distances to the Galactic center}\label{sect:Rgc}

The Arches cluster is located at a projected distance of $\sim 30$\,pc
 from the Galactic center, the Quintuplet at $\sim 35$\,pc.  These
measurements provide lower limits to the true distances of these
clusters from the center.

The mean density within the Jacobi radius of a tidally limited cluster
is proportional to the local stellar density.  (This is not a matter
of definition---it happens to be true for point-mass fields if one
averages the point mass over $\rgal$, and for the power-law density
profiles we consider here, since the tidal field $\sim M/R^3 \sim
M_G/\rgal^3 \sim \rgal^{-2}dM_G/d\rgal \simeq \rho_g$, but it is not in
general the case.)  Thus, a cluster closer to the Galactic center is
more compact, has a shorter relaxation time, and therefore evolves
more rapidly than a similar cluster at a greater distance.  

Most of our 34\,pc models have densities higher than any of the
observed clusters listed in Tab.\,\ref{Tab:observed} (see
Fig.\,\ref{fig:SPZ_W147R34N12k_TRL}).  Only the model with an
extremely shallow initial density profile (R34W1) has density
comparable to the observed systems.  Clusters at greater distances
 from the Galactic center (but with the same projected distance) have
lower densities.  Since the densities of the 34\,pc models are too
high by at least a factor of two, this suggests that the real clusters
are somewhat farther out, at $\rgal\apgt50$\,pc; a factor of 2 in
density corresponds to a factor of $2^{1/3}$ in radius, so the true
Galactocentric distance should exceed 43 pc.

\begin{figure}[htbp!]
\hspace*{1.cm}
\psfig{figure=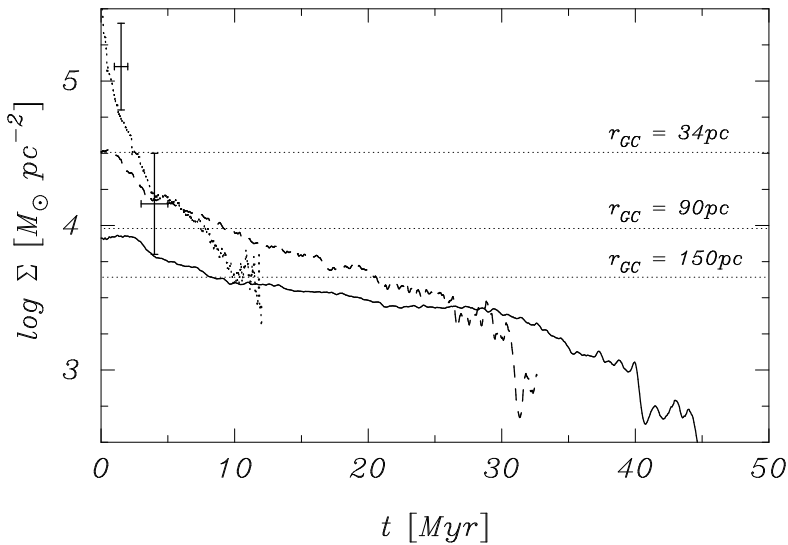,width=12.cm,angle=-90}
\caption[]{Evolution of the projected density within the half-mass
radius for models R34W7 (dots), R90W4 (dashes), and R150W1 (solid).
The dotted horizontal lines gives the surface density at a projected
distance of 34\,pc, 90\,pc and 150\,pc (as indicated) from the
Galactic center.  The selected cases are quite extreme; R34W7 being
the most concentrated model and R150W1 the least concentrated. The two
error bars (to the left) indicate the location of the Arches (left)
and the Quintuplet (right) clusters.}
\label{fig:SPZ_W17R34_rho_phm}
\end{figure}

Figure\,\ref{fig:SPZ_W17R34_rho_phm} plots the evolution of the mean
surface density within the projected half-mass radius for models R34W7
(dots), R90W4 (dashes) and R150W1 (solid), which brackets the central
densities of the models.  We view the clusters along the $x$-axis, so
we look through the second and first Lagrangian points toward the
Galactic center, giving the highest possible cluster density and hence
an overestimate of the true density contrast (see
Fig.\,\ref{fig:SPZ_W147}).  The two error bars give the projected
half-mass densities for the Arches (left), and the Quintuplet (right)
clusters.  The horizontal dotted lines give the background surface
densities a projected distances of 34\,pc, 90\,pc and 150\,pc from the
Galactic center.

\begin{figure}[htbp!]
\hspace*{1.cm}
\psfig{figure=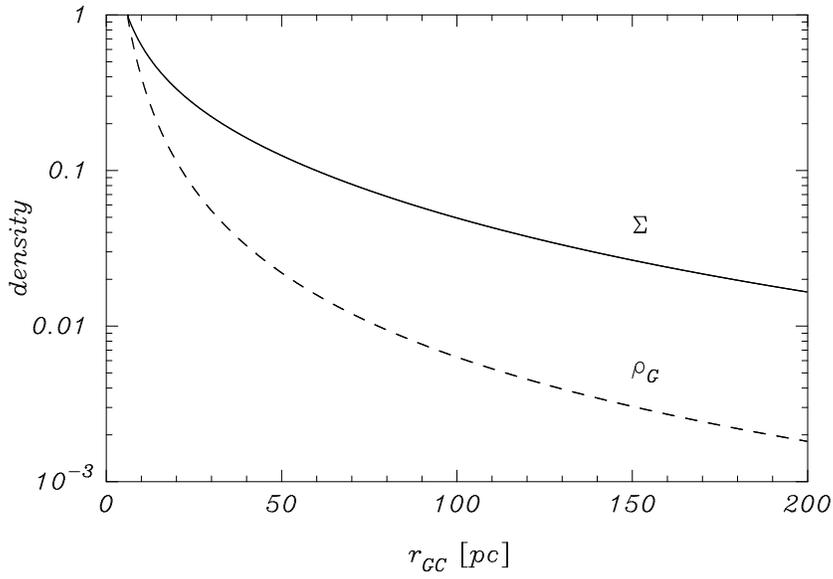,width=12.cm,angle=-90}
\caption[]{Projected (solid line, Eq.\,\ref{Eq:projectddensity}) and
three dimensional (dashed line, Eq.\,\ref{Eq:rhogal}) density as a
function of the distance to the Galactic center.  Both densities are
normalized to a distance of $\rgc=5$\,pc, }
\label{fig:density}
\end{figure}

\begin{figure}[htbp!]
\hspace*{1.cm}
\psfig{figure=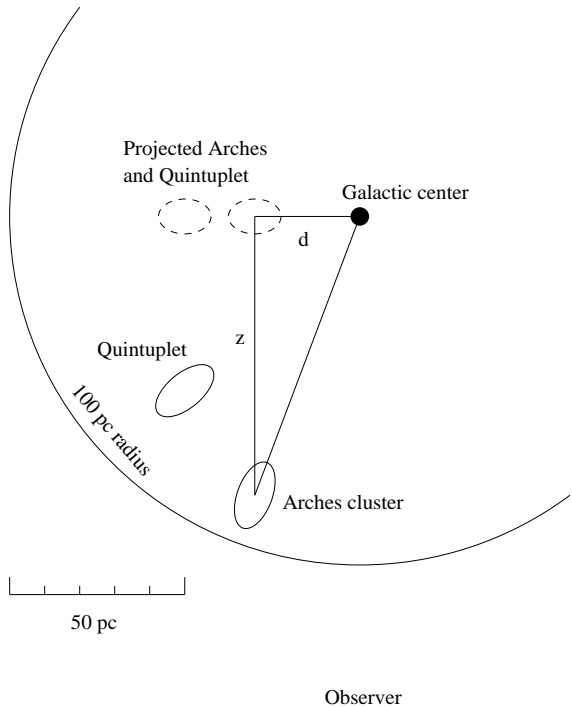,width=7.5cm,angle=-90}
\caption[]{Cartoon of the location of the Arches (left) and the
Quintuplet (far left) clusters with respect to the Galactic center.
Earth is below.  The large partial circle has a radius of 100\,pc.
The ellipses representing the clusters are exaggerated by a factor of
10.}
\label{fig:los_GC}
\end{figure}

The surface density at projected distance $d$ from the Galactic center
can be calculated by integrating the local stellar density
$\rho_G$ (Eq.\,\ref{Eq:rhogal}) along the line of sight to the
cluster:
\begin{equation} 
   \Sigma(d) = 4.06\times10^5 \int \left( d^2 + z^2 \right)^{-0.9} dz
		\;\; \msun\,{\rm pc}^{-2}.
\label{Eq:projectddensity}
\end{equation}
Integrating Eq.\,\ref{Eq:projectddensity} from -100 to 100\,pc (the
range of validity of the Mezgers' equation) gives the projected
surface densities towards the Arches (at $d=30$\,pc) and Quintuplet
cluster (at $d=35$\,pc).  We integrate Eq.\,\ref{Eq:projectddensity}
numerically.  Figures\,\ref{fig:density} and \ref{fig:los_GC}
illustrate this.  An extra correction for stars between 100\,pc from
the Galactic center and Earth adds little ($\aplt 10$\%) to the total
(Bahcall \& Soneira 1980).\nocite{1980ApJS...44...73B} The projected
densities of the observed clusters are comparable to the background.
Clusters with lower background densities may well remain unnoticed
among the background stars.  Note, however, that the background
stellar population is probably older than the star clusters studied
here, and may therefore have a smaller mass-to-light ratio, making the
clusters stand out somewhat better.  This may be the reason that the
Quintuplet was found even though its projected density is smaller than
that of the background.

Figure\,\ref{fig:density} shows Eq.\,\ref{Eq:projectddensity} and
Eq.\,\ref{Eq:rhogal} both normalized to their value at $\rgc=5$\,pc,
which are $\Sigma(\rgc=5{\rm pc}) = 1.6\times 10^5$\,\msun\,pc$^{-2}$
and $\rho_G(\rgc=5{\rm pc}) = 1.6\times 10^4$\,\msun\,pc$^{-3}$.

The projected density of model R150W1 remains well below the
background, and such a cluster could easily remain unseen throughout
its entire lifetime.  The two initially more concentrated models,
R90W4 and R34W7, have projected densities above the background, at
least initially.  The cluster farther from the Galactic center has a
lower initial density because it is more extended; after the first few
million years, it may become hard to see.  The observation of the
Quintuplet provides an upper limit on the critical contrast below
which the cluster cannot be detected.  We arbitrarily adopt a minimum
contrast equals to the surface density of the Quintuplet cluster
(i.e. $\sim 10^4$\,\msun\,pc$^{-2}$) as a threshold for distinguishing
a star cluster from the background.  In that case, the 150\,pc cluster
would be invisible for its entire lifetime, while the 34\,pc and the
90\,pc clusters remain visible for about $\sim 8$\, million years
($\sim 65$\% and $\aplt 25$\% of their respective lifetimes).  Thus,
although the clusters farther from the Galactic center live much
longer, their visible lifetime is actually less than that of a cluster
at smaller Galactocentric radius.

Although not shown in Fig.\,\ref{fig:SPZ_W17R34_rho_phm}, the surface
density evolution of model R34W1 is also consistent with the densities
of the observed clusters.  On these grounds we therefore cannot
exclude the possibility that the Arches and Quintuplet systems lie at
Galactocentric distances of 34\,pc, but in that case they must have
been born with very shallow density profiles (which, however, would
contradict our earlier analysis from Sec.\,\ref{sect:mass_function},
see Tab.\,\ref{Tab:Arches}).

The surface mass density as presented in
Fig.\,\ref{fig:SPZ_W17R34_rho_phm} may not be the ideal way to express
the visibility of these clusters; a luminosity density contrasted with
the background would be more appropriate. This, however, requires a
detailed study of the observational selection effects, which goes
beyond the scope of this paper. Another way to illustrate the
visibility of clusters like Arches is by looking at the integrated
visual luminosity of these clusters (see Fig.\,\ref{fig:TL_W4150}).

\section{Discussion}\label{Sect:Discussion} 

The Arches and the Quintuplet are among the youngest star clusters
known in the Galaxy.  They are not as rich as small globular clusters,
but are considerably more compact.  Their central densities are
comparable to those of the densest---post collapse---globulars, but
their life expectancies are only a few tens of millions of years.
These clusters are destroyed by tidal forces, accelerated by impulsive
mass loss from supernovae (after $\sim3$ Myr) and strong binary
activity.  In the following we first focus on comparison with other
calculations; we then discuss the consequences of our results for the
inferred distances of the Arches and Quintuplet clusters from the
Galactic center, and for the number of such clusters which may be
hidden near the Galactic center.

\subsection{Comparison with other work}\label{sect:comparison}
Portegies Zwart et al.~(1999, \paperIII)\nocite{pzmmh99} studied the
evolution of R\,136 in the LMC using {\nbody} simulations.  Their
initial conditions were comparable to those adopted here, although
small differences exist.  They adopted King models with $\Wo=6$,
whereas we selected $\Wo=4$ and $\Wo=7$.

The calculations in {\paperIII} included the effects of stellar
evolution and physical collisions between stars, but neglected the
effect of the Galactic tidal field and the possible presence of
primordial binaries.  Both may have a profound effect on a cluster's
dynamical evolution.  For R\,136 the neglect of a tidal field may be
appropriate, but for the Arches and Quintuplet clusters the tidal
field is crucial.  Our clusters therefore dissolve much more
rapidly than the calculations in {\paperIII}.  The latter models lost
only a few percent of their mass within the first relaxation time,
where our models lose up to $\sim 50$\% on this time scale, confirming
the strong influence of the Galactic tidal field.

Portegies Zwart et al.~(1999) found that physical
collisions between stars in their models were frequent, and that the
evolution of the most massive stars and the dynamical evolution of the
cluster were closely coupled.  In all cases, a single star grew
steadily in mass through mergers with other stars, forming a very
massive ($\apgt100\msun$) star in less than 3--4 Myr.  The growth rate
of this runaway merger was much larger than estimates based on simple
cross-section arguments, mainly because the star was typically found
in the core and tended to form binaries with other massive stars
there.  We observe the same general behavior in our calculations, in
the sense that collisions tend to occur between high-mass
($m\apgt10$\,\msun) stars and that most collisions occur repeatedly
with the same star. 
Although not discussed in detail in this paper, we
find the same trend in our calculations and confirm the findings of
Portegies Zwart et al.~(1999).
Kim, Morris \& Lee (1999; KML99)\nocite{1999ApJ...525..228K} performed
2-D {\FP} calculations of dense star clusters near the Galactic center
and studied the lifetimes of the Arches and the Quintuplet clusters.
Their calculations are somewhat different than those presented in this
paper in the sense that they only solve the two dimensional {\FP}
equation instead of the equation of motions for all stars in the
stellar system.  KML99 include standard binary heating (Lee et al.\,
1991)\nocite{1991ApJ...366..455L} in their calculations and neglect
stellar collisions.

\begin{figure}[htbp!]
\hspace*{1.cm}
\psfig{figure=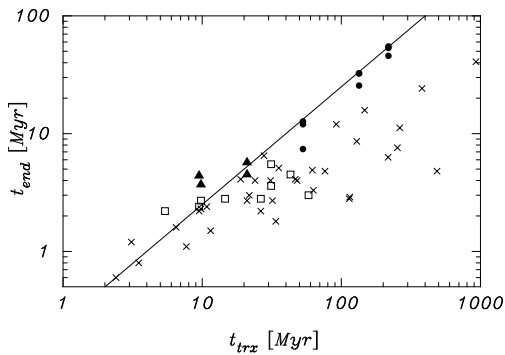,width=7.5cm,angle=-90}
\caption[]{Cluster lifetime as a function of the initial relaxation
time at the tidal radius for the model calculations presented in
table.\,\ref{Tab:N12kinit} ($\bullet$, 9 models), those presented in
Kim, Morris \& Lee ($\times$) and in Kim et al (2000; open squares).
The solid line ($\tend = 0.25\trxt$) presents the scaling of the
models as proposed by Portegies Zwart et al.\, (2001a). The triangles
are calculated as an extra check on the behavior of the models when
$N\aplt 3000$, which represents the parameter space covered by Kim et
al.\, (2000; their models 101 111 112 and 141).}
\label{fig:PZvsKea}
\end{figure}

We selected the most representative cases of KML99 and recalculated
these with our model.  The Fokker Planck models dissolve much more
rapidly (often more than a factor two in lifetime) than the \nbody\,
models (see Fig.\,\ref{fig:PZvsKea}).  Similar discrepancies are
observed in the comparison between the Fokker-Planck calculations of
Chernoff \& Weinberg (1990)\nocite{1990ApJ...351..121C} and the
\nbody\, calculations of Fukushige \& Heggie
(1995)\nocite{1995MNRAS.276..206F} and Portegies Zwart et al.\,
(1998).\nocite{1998A&A...337..363P} The discrepancies are the result
of the rather simple treatment of escapers in the Fokker-Planck
models, which can drive the evaporation of the cluster on a shorter
time scale then when a self-consistent tidal field is used (see
Takahashi \& Portegies Zwart 1998\nocite{1998ApJ...503L..49T} for
details). An extra complication is introduced in the calculations of
KML99 by selecting a maximum mass of 150\,\msun\, in the initial mass
function. Such stars dominate the early evolution of the star cluster
while their behavior cannot accurately be represented with a
statistical average as is done in a Fokker-Planck solver.  Part of the
spread in the Fokker-Planck results in Fig.\,\ref{fig:PZvsKea} may be
attributed to these effects.

Kim et al. (2000)\nocite{2000ApJ...545..301K} also perform N-body
calculations of several of the models explored by KML99 using
Aarseths' (1999)\nocite{1999PASP..111.1333A} {\tt NBODY6}.  They also
find that the N-body models live longer than the Fokker-Planck models,
but arrive at the same conclusion as KML99: the Arches and Quintuplet
star clusters must have unusually flat initial mass function and
dissolve within about 10\,Myr.  The results of KML99 are completely
understandable from their choice of initial conditions.  Many of their
calculations start with a Power-law initial mass function with and
exponent as flat as -1.5 (Salpeter $= -2.35$) and omitting all stars
with a mass smaller than 1\,\msun. Together with a small number of
stars of only a few thousand, their models have an initial
relaxation time at the tidal radius of only $\trxt=5$ to 60\,Myr. The
lifetime of these clusters is therefore expected to be only a few
million years.

Figure\,\ref{fig:PZvsKea} compares the results of our calculations
($\bullet$ and triangles) with the Fokker-Plank calculations of KML99
($\times$), the N-body calculations of Kim et al. (2000, squares).
For comparison we added the results of the analytic model presented by
Portegies Zwart et al.\, (2001a) as a solid line.  The lowest values
of our models (lower $\bullet$ for each \trxt) are computed with
$\Wo=1$, which dissolve somewhat quicker than the more concentrated
models.  The lifetime of the {\nbody} models start to depend on $N$
via the Coulomb logarithm (see Eq.\,\ref{Eq:trlx}) when the number of
stars in the models drops below a few thousand.  In this small $N$
limit the scaling proposed by Portegies Zwart et al (2001a) probably
breaks down. This effect may cause our models with the shortest
initial relaxation times to live longer than expected (leftmost
triangles in Fig.\,\ref{fig:PZvsKea}).  A similar effect may cause the
models of Kim et al.\,(2000, open squares) to have somewhat shorter
lifetimes than expected, as most of their models contain less than
3000 stars.  Part of the discrepancy at small $N$ and therefore at
small relaxation times may also be caused by statistical fluctuations,
which become most noticeable at small $N$.

\subsection{How many clusters are still hidden near the Galactic
center}\label{Sect:visibility}  
Although the Arches and Quintuplet clusters are very dense, it may
still be difficult to distinguish them near the Galactic center.  The
three-dimensional density contrast is several orders of magnitude, as
can be seen in Fig.\,\ref{fig:SPZ_W147R34N12k_TRL}.  In projection on
the sky, however, the density contrast is reduced, due mainly to the
accumulation of stars along the line of sight (see
Fig.\,\ref{fig:SPZ_W17R34_rho_phm}).  The cluster density within the
projected half-mass radius also decreases with time, so these clusters
are more easily seen at early ages than at later times.  As outlined
in Sec.\,\ref{sect:Rgc}, to estimate the time over which our model
clusters would be observable, we simply compare the integrated stellar
density $\Sigma(d)$ along the line of sight to the cluster
(Eq.\,\ref{Eq:projectddensity}) with the projected stellar densities
of the models.

In section\,\ref{sect:Rgc} we introduced a limiting density contrast
above which a cluster can be discriminated from the background stars,
and used this to estimate the time during which our models would be
detectable.  At a Galactocentric distance of 34\,pc, our most compact
models (R34W7) remain visible for $\sim9$\,Myr; at 90\,pc they become
invisible after about 7\,,Myr and at a distance of 150\,pc from the
Galactic center they remain invisible for their entire lifetime.
Models at greater distances from the Galactic center, or those born
with shallower density profiles, remain undetectable for their entire
life times.  It is therefore not surprising that the clusters we
observe near the Galactic center are extremely compact and very young.
Less compact or older clusters are unobservable due to their low
surface density contrasts.


Portegies Zwart et al.\, (2001a)\nocite{2001ApJ...546L.101P} studied
the time scale over which clusters such as Arches and the Quintuplet
systems remain visible.  The results of their detailed N-body
calculations are used to calibrate a simple analytical model which is
applicable over a wider range of cluster initial conditions.  They
conclude that clusters within 200\,pc of the Galactic center dissolve
within $\sim70$ Myr.  However, their projected densities drop below
the background density in the direction of the Galactic center within
$\sim 20$\,Myr, effectively making these clusters undetectable after
that time.  Clusters farther from the Galactic center but at the same
projected distance are more strongly affected by this selection
effect, and may go undetected for their entire lifetimes.  Based on
these findings, they conclude that the region within 200 pc of the
Galactic center could easily harbor some 50 clusters with properties
similar to those of the Arches or the Quintuplet systems.  The results
of our more extended parameter study is consistent with their
findings. Given the higher mass we derive for the Arches cluster (see
Tab.\,\ref{Tab:Arches}) compared to what was adopted by Portegies
Zwart et al (2001a) we argue that their results may even be somewhat
conservative.

Another effect which may contribute to the difficulty in detecting
clusters like the Arches is the enormous range in luminosity of the
brightest stars.  CCD cameras have a dynamic range of less than
$2^{16}$ = 64k, causing the brightest stars to saturate the detector
and preventing faint stars from being detected.  However, these
brightest stars are also the least common; only with very deep
exposures is the rest of the cluster revealed.  Thus, wherever two or
more bright blue stars are seen together, there may be an entire star
cluster lurking in the background. An example of such a cluster may be
the small group of stars identified as R\,140\footnote{R\,140 contains
at least two WN stars and one WC star (Moffat
1987)\nocite{1987ApJ...312..612M} located at a projected distance of
11.5\,pc from R\,136. One of the WN stars (R\,140a) is a bright $\log
L_{\rm 0.2-3.5 keV} = 10^{35.2}$ erg\,s$^{-1}$ X-ray source, possibly a
colliding wind binary (Portegies Zwart et
al. 2001c).}.

Figure\,\ref{fig:TL_W4150} presented the integrated luminosity for two
representative models (R90W4 and R150W4). In these models the total
luminosity dropped by more than two orders of magnitude over the
lifespan of the cluster. The individual high peaks are the results of
single stars which become extremely bright as a result of their
rejuvenation in several collisions. Such stars can temporarily
outshine the rest of the star cluster.  Of course, the luminosity and
the lifetime of such a multiple collision product is quite sensitive
to details in modeling the collision process.

\section{Conclusion}\label{Sect:Conclusion}
We have studied the evolution of young, dense star clusters near the
Galactic center, taking the Arches and Quintuplet clusters as specific
examples. These clusters are generally referred to as ``young,''
because they are only a few million years old, and indeed, even in a
dynamical sense both clusters are quite young, having lifetimes
considerably smaller than their initial relaxation times. However, by
observing only the most massive stars in these clusters the dynamical
image sketched by the observers is biased toward greater age, as the
massive stars evolve dynamically on shorter time scales than average
cluster members.  By observing only stars with $m\apgt10\,\msun$, one
selects a part of the cluster that is dynamically rather mature, being
comparable in age to the local initial relaxation time.  This makes
the Arches cluster (and probably also the Quintuplet) appear
dynamically older than it really is.  We cannot test this hypothesis
for the Quintuplet system because the available data are of lower
quality than for the Arches.

The modeled clusters lose mass at a more-or-less constant rate,
inversely proportional to the initial relaxation time, i.e.\, $\propto
\rhm^{-3/2} \propto \rgc^{-0.9}$ (see also Portegies Zwart et al.
2001a).  Star clusters which are born farther from the Galactic center
live therefore considerably longer than those closer in.  The
relaxation time derived from observation, however, contains little
information about the initial relaxation time.

Mass segregation in our model clusters quickly causes the most massive
stars to sink to the cluster center. As a result core collapse occurs
within about two million years, even for models as shallow as
\Wo=1. During core collapse close binaries are formed and collisions
between stars are frequent.  The collision rate is much higher than
would be expected from simple cross section arguments, due to mass
segregation and binary formation. We confirm the finding of Portegies
Zwart et al. (1999)\nocite{pzmmh99} that the most massive stars are
generally involved in a collision runaway, in which few low mass stars
participate. This process continues until the runaway product is
ejected by a supernova or a strong encounter with a binary or the
cluster dissolves in the tidal field of the Galaxy.

By comparing mass functions from our models at various distances from
the cluster center with observations of the Arches mass function, we
conclude that the age of the Arches cluster must be about 0.05\,\trxh.
With an age of about 1.5\,Myr the initial relaxation time of the
cluster is then about 30\,Myr. To reconcile the observed mass function
with the observed density distribution of the Arches, we conclude that
the cluster must be about $4 \times 10^4$\,\msun\, and lie some 50 --
90\,pc from the Galactic center. With these parameters we can
reproduce the observed density distribution and the observed mass
function for stars between 10\,\msun\, and 50\,\msun\, without
requiring an unusually flat initial mass function.  The top end of the
mass function contains more stars than our adopted mass function and
this may require a flatter $~\aplt -2.8$ (Salpeter $= -2.35$) initial
mass function for stars $\apgt 50$\,\msun.  However, we still conclude
that the Arches cluster can be explained with a ``normal'' mass
function. Most of the observed flattening of the mass function can be
attributed to selection effects caused by the limited range in
distance from the cluster center in which the observations were made.

The young clusters we discussed expand as they become older, causing
their surface densities and total luminosity to drop.  Clusters which
are older than about 5\,Myr often have surface densities below that of
their surrounding, making these clusters virtually undetectable.  By
comparing the projected density of the model clusters with that of the
field stars in the direction of the Galactic center we conclude that
the region within 200 pc of the Galactic center could easily harbor
some 50 clusters with properties similar to those of the Arches or the
Quintuplet systems.  To find more clusters like Arches and Quintuplet
systems, we urge observers to look for single extremely bright stars.
Closer study of the region near such a star may well reveal the
underlying star cluster.

Finally, with our estimate of about 50 clusters with a mass of $\sim
40\,000$\,{\msun} each and dissolving within 100\,Myr, we derive a
star formation rate near the Galactic center of
$\sim0.02$\,\msunyr. Clusters like the Arches and Quintuplet may
therefore contribute significantly to the total star formation rate in
the Galaxy.

\acknowledgements We thank Rob Olling and Mark Morris for discussions,
and the excellent comments of the anonymous referee.  This work was
supported by NASA through Hubble Fellowship grant HF-01112.01-98A
awarded by the Space Telescope Science Institute, by ATP grants
NAG5-6964 and NAG5-9264, and by the Research for the Future Program of
Japan Society for the Promotion of Science (JSPS-RFTP97P01102).  SPZ
is grateful to Drexel University, Tokyo University and to the
Institute for advanced study for their hospitality and the use of
their fabulous hardware.  Part of this letter was written while we
were visiting the American Museum of Natural History.  We acknowledge
the hospitality of their astrophysics department and visualization
group.  We also acknowledge the expert visualization help offered to
us by Stuart Levy, from the Virtual Director Group, National Center
for Supercomputing Applications, University of Illinois at
Urbana-Champaign.  We thank the Alfred P. Sloan Foundation for a grant
to Hut for observing astrophysical computer simulations in the Hayden
Planetarium at the Museum.

\end{document}